\documentclass[12pt]{article}

\usepackage{scicite}
\usepackage{graphicx}
\usepackage{subcaption}
\usepackage{multirow}
\usepackage{booktabs}
\usepackage{bm}
\usepackage{lineno}
\usepackage{upgreek}

\usepackage{makecell}

\usepackage{times}
\usepackage{sidecap}
\usepackage{amsmath}
\usepackage{amssymb}

\usepackage{tikz}

\newenvironment{sciabstract}{%
\begin{quote} \bf}
{\end{quote}}

\title{Title: Tomography of ultra-relativistic nuclei with polarized photon-gluon collisions\\
{\normalsize{ Short Title: Quantum Interference Enabled Nuclear Tomography}} }

\author{STAR Collaboration}

\date{}

\begin{document} 



\maketitle 

\begin{sciabstract}
A linearly polarized photon can be quantized from the Lorentz-boosted electromagnetic field of a nucleus traveling at ultra-relativistic speed.
When two relativistic heavy nuclei pass one another at a distance of a few nuclear radii, the photon from one nucleus may interact through a virtual quark-antiquark pair with gluons from the other nucleus forming a short-lived vector meson (e.g. $\boldsymbol{\rho^0}$). In this experiment, 
the polarization was utilized in diffractive photoproduction to observe a unique spin interference pattern in the angular distribution of $\boldsymbol{\rho^0\rightarrow\pi^+\pi^-}$ decays. The observed interference is a result of an overlap of two  wave functions at a distance an order of magnitude larger than the $\boldsymbol{\rho^0}$ travel distance within its lifetime. The strong-interaction nuclear radii were extracted from these diffractive interactions, and found to be $\boldsymbol {6.53\pm 0.06$ ${\rm fm}}$ ($\boldsymbol {^{197} {\rm Au }}$) and $\boldsymbol {7.29\pm 0.08$ ${\rm fm}}$ ($\boldsymbol {^{238} {\rm U}}$), larger than the nuclear charge radii. 
The observable is demonstrated to be sensitive to the nuclear geometry and quantum interference of non-identical particles.
\end{sciabstract}


%
\baselineskip24pt

\section*{TEASER}
Polarized photon-gluon fusion reveals quantum wave interference of non-identical particles and shape of high-energy nuclei
\section*{INTRODUCTION}
Gluons are mediators of the strong interaction, the force that binds quarks inside protons and neutrons. 
Similar to electromagnetic interactions that occur between objects carrying electric charge, strong interactions occur between particles that carry the strong charge, color. 
It is the exchange of color between quarks and gluons that prevents them from existing as free particles and makes them inseparable from the nuclear core. 
Despite the success of Quantum Chromodynamics (QCD), the theoretical framework that describes the nature of the strong interaction, many quantities cannot be computed with existing methods, making several questions about the nature of gluons inside nuclear matter inaccessible.  
For example, current QCD calculations cannot describe the momentum-dependent parton distribution functions (PDFs) of gluons inside the proton, or explain why, at certain energy and momentum scales, gluons are not confined within the bounds of the protons and neutrons that make up a nucleus. 
Mapping out the dynamical distribution of gluons inside nuclei, which determine the nuclear size at high energy, and understanding how their interactions produce the vast majority of the mass of the visible universe has become a major driving force in modern experimental nuclear science~\cite{EICAccardi:2012qut}.

\begin{figure}
    \centering    
    \includegraphics[width=0.75\textwidth]{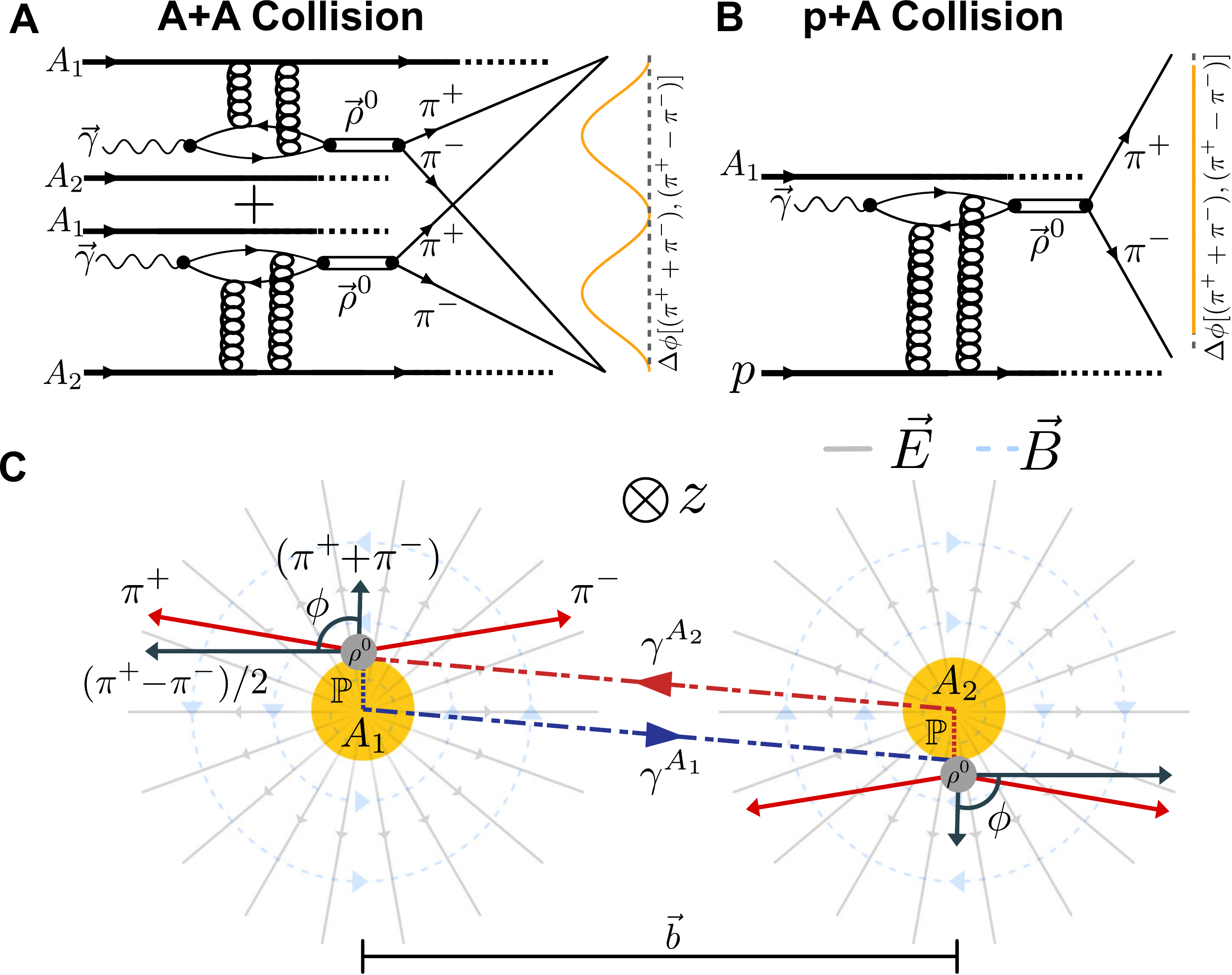}
    \caption{ \label{fig:fig_1} (A) A Feynman-like diagram for a gold-gold interaction in which there is exclusive photo-nuclear production of a $\rho^0$ meson that subsequently decays to a $\pi^+\pi^-$ pair. Quantum interference between the transverse linear polarization from the photon in each diagram results in an observed $\cos2\phi$ dependence despite the two diagrams not sharing any internal lines. 
    (B) A diagram for the same process in a proton on gold interaction, where essentially no interference takes place due to the large difference in charge between the proton and the gold nucleus.
    (C) An illustration of a photo-nuclear interaction occurring between two ultra-relativistic nuclei separated by a nucleus-nucleus impact parameter ($\vec{b}$) of several nuclear radii. While only one $\rho^0$ is produced, two possible configurations contribute to the amplitude, one where a photon is emitted by the field of nucleus 1 ($A_1$) and a Pomeron by nucleus 2 ($A_2$), and vice-versa. Vectors representing the 2D momentum in the plane transverse to the beam (along the z-axis) are shown for the photons ($\gamma$), Pomerons ($\mathbb{P}$), $\rho^0$, and $\pi^\pm$. The angle $\phi$, a proxy for the $\rho^0$ polarization direction, is defined in terms of the sum and difference of the 2D momentum of the $\pi^+$ and $\pi^-$.  }
\end{figure}

Just as visible light in quantum optical coherent tomography~\cite{Graciano_2019} and coherent X-rays are used to study biosamples, cells and atoms, high-energy photons can be used to probe gluons inside of nuclei.  While photons do not interact directly with gluons because they don’t carry color, interactions can still occur when the photon temporarily fluctuates into a quark-anti-quark pair that in turn interacts with the gluons inside the nucleus. The simplest diagram for this type of interaction is shown in Fig.~\ref{fig:fig_1}A for nucleus-nucleus ($A+A$) collisions and in Fig.~\ref{fig:fig_1}B for proton-nucleus ($p+A$) collisions. In these diagrams a photon interacts with a Pomeron, a color-neutral two-gluon state at lowest order, and produces a vector meson ($\rho,\phi,J/\psi$, etc.)~\cite{star_collaboration_coherent_2017,ALICE:JPSI2021tyx,STAR:JPSI2021wwq} which has the same intrinsic quantum numbers as the incoming photon. Besides Pomerons, other configurations such as Reggeons involving quarks may contribute to the scattering at lower energies~\cite{LEBIEDOWICZ2014301,Alvensleben:1970uw,CornellPhysRevD.4.2683}. 
However, at the energies investigated here, interactions occur primarily via Pomerons and are therefore sensitive to the gluon distribution of the colliding nuclei~\cite{starcollaborationMeasurementCentralExclusive2020}.
For this reason, exclusive photoproduction of a vector meson, i.e. when the two outgoing nuclei remain intact, provides a unique opportunity to probe the gluonic structure of nuclear matter. 
This process applied to nuclear matter is analogous to positron emission tomography (PET) scanning originally developed for imaging of the human brain~\cite{PET1953}. 
Just as PET scanning, first proposed in the 1950s, has undergone decades of technological evolution~\cite{jonesDevelopmentAchievementsFuture2012,PET2013Portnow952} resulting in the precision tool that we have today, this gluon tomography technique, achieved here for the first time, provides the closest technology to the 3D gluon imaging planned at the future electron-ion collider~\cite{EICAccardi:2012qut}.


Although the proton PDFs have been studied extensively in high-energy $e$+$p$ collisions at the Hadron Electron Ring Accelerator (HERA) and other facilities~\cite{pdg2020} (see section 18 on structure functions and references therein), currently there is no $e$+$A$ collider for such a study of nuclear matter. Even without an $e$+$A$ collider, similar measurements have been performed in ultra-peripheral $A$+$A$ collisions (UPC), where the nuclei pass with a nucleus-nucleus impact parameter ($b$) large enough to avoid nuclear contact. 
The equivalent incident photon energy, in the rest frame of the target nuclei, is in the range of 50-100 GeV at Relativistic Heavy Ion Collider (RHIC) and is an order of magnitude higher at the Large Hadron Collider (LHC). Indeed, measurements at RHIC and LHC have employed these high-energy photons to study nuclear structure and the gluon distribution within nuclei at high energy~\cite{star_collaboration_coherent_2017,alicecollaborationCoherentPhotoproductionRho2020a} in the past. 
However, there remain several open questions~\cite{Ryskin:1992ui,
Klein:1999qj,Munier:2001nr, 
Guzey:2016piu
} and uncertainties hindering the extraction of the gluon density distribution at a quantitative level. 
Among the most problematic are the uncertainty of the photon source generated by the Coulomb field of the heavy nuclei, the separation of coherent diffractive production from incoherent production, and a method for modeling the process that matches the observed data with high precision. 
As we will show in this Article, these issues and others have so far made precision measurement of the gluon distribution of nuclei at high energy unreliable, in some cases leading to an apparent mass radius more than $1$~fm larger than the charge radius\cite{star_collaboration_coherent_2017}. 

Previously, it was perceived that photons participating in UPC events are quasi-real with transverse momentum $k_\perp~=1/R_p\sim30$ MeV (note that natural units are used throughout), where $R_p$ is the nuclear charge radius, reflecting the virtuality and uncertainty principle of their origin. 
This led to the assumptions in models employing the equivalent photon approximation (EPA)~\cite{Baltz:2007kq,Klein:2018cjh,Klein:2018fmp,Schafer:2020bnm} that processes initiated by these photons do not depend on the nucleus-nucleus impact parameter and that their transverse spatial coordinates are randomly distributed based on the same principles. 
The recent measurements of lepton pair production from photon collisions in UPC events at RHIC~\cite{starcollaborationProductionEnsuremathPairs2004,Afanasiev:2009hy,PhysRevLett.121.132301,starcollaborationMeasurementMomentumAngular2021} and LHC~\cite{Abbas:2013oua,CMS:2020skx,Acharya:2018nxm,Aaboud:2018eph} have shown that the photons behave as real photons in all observables.
More importantly, these measurements provide a precision calibration~\cite{Li:2019yzy,Klein:2020jom} necessary for the photons to be a source for the photo-nuclear processes, and therefore gluon tomography.

Instead of being randomly distributed spatially, and therefore randomly polarized, recent measurements from the STAR experiment~\cite{starcollaborationMeasurementMomentumAngular2021} demonstrate that the quasi-real photons participating in photon-photon and photo-nuclear $A+A$ collisions are linearly polarized in the transverse plane.
In photo-nuclear interactions, the linear polarization of the spin-1 photon is transferred, without a helicity flip, to the produced vector meson~\cite{theh1collaborationDiffractiveElectroproductionRho2010,DERRICK1996259}. Upon decay, the spin in the system is transferred into the orbital angular momentum (OAM) of the daughter particles. 
Such spin information has previously been inaccessible though, since the photon polarization, oriented with the nucleus-nucleus impact parameter ($\vec{b}$), is random from one event to the next. On the other hand, recent theory calculations referred to as Model I~\cite{Zha:2018jin,ZhaPhysRevD.103.033007} and Model II~\cite{Xing:2020hwh} have shown that a $\cos 2\phi$ asymmetry exists due to the linear polarization of the incident photons, where $\phi$ is the angle in the transverse plane between the vector meson's momentum and one of the daughter's momentum (see Eq.~\ref{eq:phiangle}). Since the daughters' momenta are used as a proxy for the $\rho^0$ spin direction, the accuracy of such a measure has a resolution of $\epsilon_p=\langle\cos{2\phi}\rangle=1/2$ for linear polarization, similar to the reaction plane resolution in elliptic flow analyses~\cite{STAR:2000ekf}, which contributed to the discovery of the strongly interacting quark-gluon plasma~\cite{STARQGPWP:2005gfr}.

Unlike past theoretical models, both of these models (Model I and Model II) implement a correlation between the incoming photon's spin and momentum. Model I implements a photon and Pomeron interaction using a Woods-Saxon distribution with the $\rho^0N$ cross section~\cite{ZhaPhysRevD.103.033007}. Model II implements a dipole and gluon interaction with the gluon distribution inside the nucleus given by a Color-Glass Condensate (CGC) model including the $\rho^0$ wave function contribution~\cite{Xing:2020hwh}. A more detailed discussion of the models is given in later sections. 
Crucially, both models predict that the alignment in the final-state between the vector meson's momentum and the momenta of its daughters is expected to result from interference between the contributing amplitudes shown in Fig.~\ref{fig:fig_1}A. 
We emphasize that in this experiment, the final observable is only one $\pi^+\pi^-$ pair and the interference of the two $\rho^0$ occurs at the wave function level with only one $\rho^0$ physically produced. 

As illustrated in Fig.~\ref{fig:fig_1}C, the wave functions of the $\rho^0$ are created at a distance on average about the impact parameter ($\langle b \rangle\simeq 20$ fm~\cite{Klein:2002gc}) apart while the lifetime of the $\rho^0$ is only about $1$ fm. This is a classic example of the Einstein–Podolsky–Rosen (EPR)~\cite{Klein:1999qj,Klein:2002gc,starcollaborationObservationTwoSourceInterference2009,Bertulani:2008gq} paradox, in which the daughter pions from the $\rho^0$ decay are assumed to maintain the overall wave function of their parent, which is required for the interference to happen. 
Multiphoton interference and entanglement have been used in many fields of fundamental research and have many applications~\cite{QERevModPhys.84.777}, even OAM interferometry with twisted light~\cite{magana-loaizaHanburyBrownTwiss}. 
In photoproduction of vector mesons off nuclei, the transition from a photon to a vector meson has been treated with the vector dominant process, i.e. by treating a vector meson like a heavy photon~\cite{schillingAnalysisVectormesonProduction1970,schillingHowAnalyseVectormeson1973a}. 
Therefore, applying interferometry in this case in analogy to photons is quite natural. 

Both Model I and Model II indicate that the transverse-momentum-dependent $\cos 2\phi$ asymmetry should have a distinctive diffractive pattern which is sensitive to the nuclear geometry and quantum interference effects. 
Interference was previously believed to be present and observable only at extremely low four-momentum transfer between the photon and the entire nucleus ($|t|$) because of a presumed random photon polarization~\cite{Klein:1999gv,Hencken:2005hb,Abelev:2008ew,Zha:2018jin,ZhaPhysRevD.103.033007,Xing:2020hwh}. 
Destructive interference results between the two participating amplitudes at very low $|t|$ due to the identical particle symmetry and the odd parity of the $\rho^0$ vector meson, causing a dip in the cross section near $|t|=0$.
In addition, this newly observed interference effect alters the $|t|$ distribution of the diffractive cross section even at larger values of $|t|$. As we will demonstrate, this modification of the $|t|$ spectra, especially at higher values of $|t|$ is primarily responsible for the anomalously large and unreliable nuclear radii extracted from past measurements of $|t|$ spectra from photo-nuclear processes.

Given that the photons are 100\% linearly polarized along their transverse momentum direction, the $\rho^0$ momentum perpendicular to the polarization has a minimal contribution from the photon transverse momentum in the $|t|$ distribution - and is therefore most indicative of the Pomeron momentum contribution. 
The interference occurs with a term $\cos{(\vec{P}\cdot\vec{b})}$ where $\vec{P}$ is the momentum vector of the $\rho^0$ and $\vec{b}$ is the nucleus-nucleus impact parameter. 
For a point source or in the large $|\vec{b}|$ limit~\cite{Zha:2018jin,ZhaPhysRevD.103.033007}, the photon transverse momentum and polarization vectors are exactly along the impact parameter direction. 
In the current work, we study the $|t|$ distribution as a function of $\rho^0$ polarization angle $\phi$. At $\phi=90^{\circ}$, all the effects from photon momentum, polarization and interference should be at a minimum or completely disappear.  Even with those contributions removed, extracting the true nuclear mass form factor requires a small correction for the $\rho$ wavefunction ($R_T^{\rho}\simeq1$ fm~\cite{theh1collaborationDiffractiveElectroproductionRho2010,Forshaw:2010py}) and the depolarization due to the finite angle between the photon polarization vector and $\vec{b}$ as depicted in Fig.~\ref{fig:fig_1}C.

\begin{figure}
    \centering    
    \includegraphics[width=0.49\textwidth]{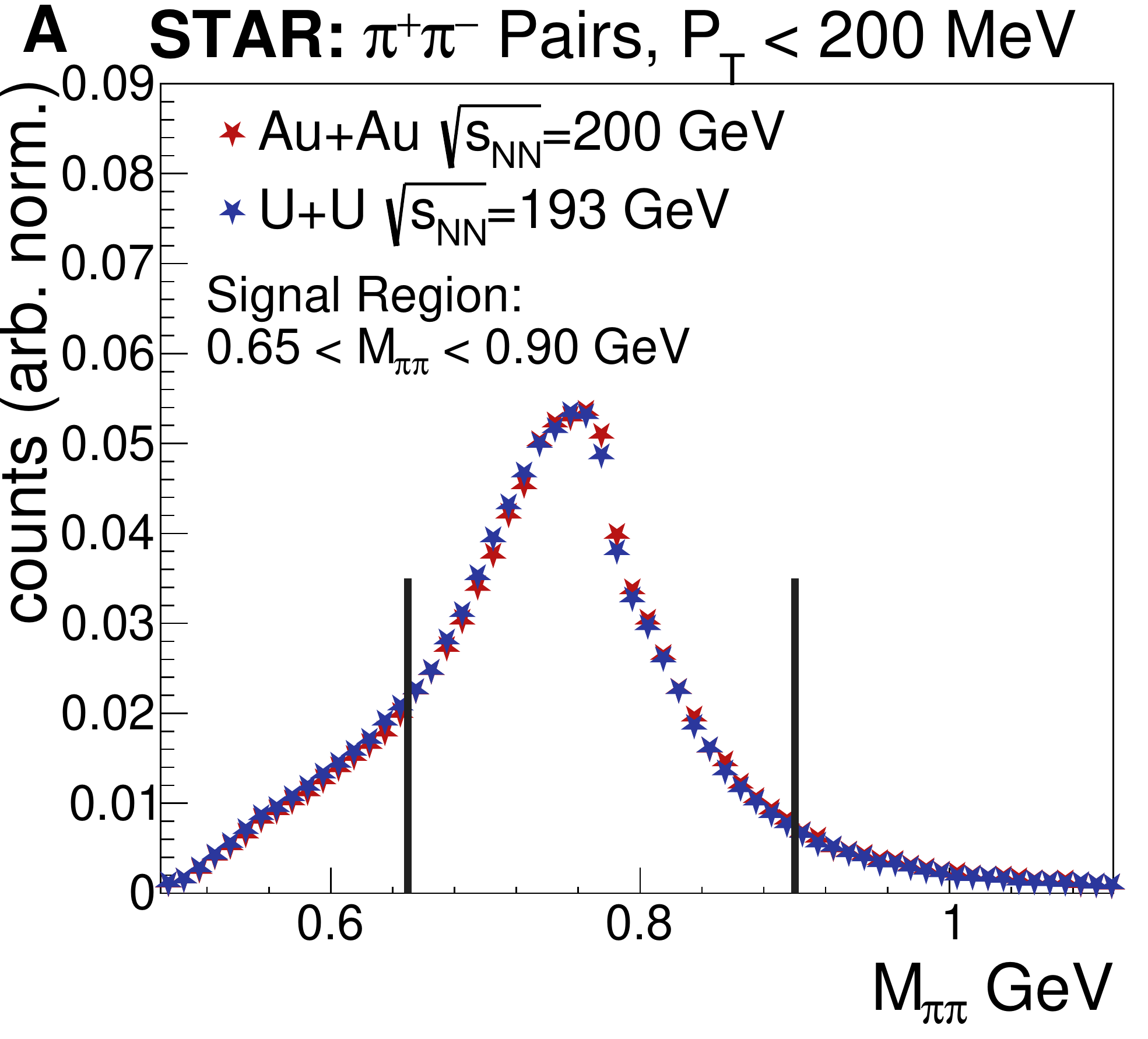} 
    \includegraphics[width=0.49\textwidth]{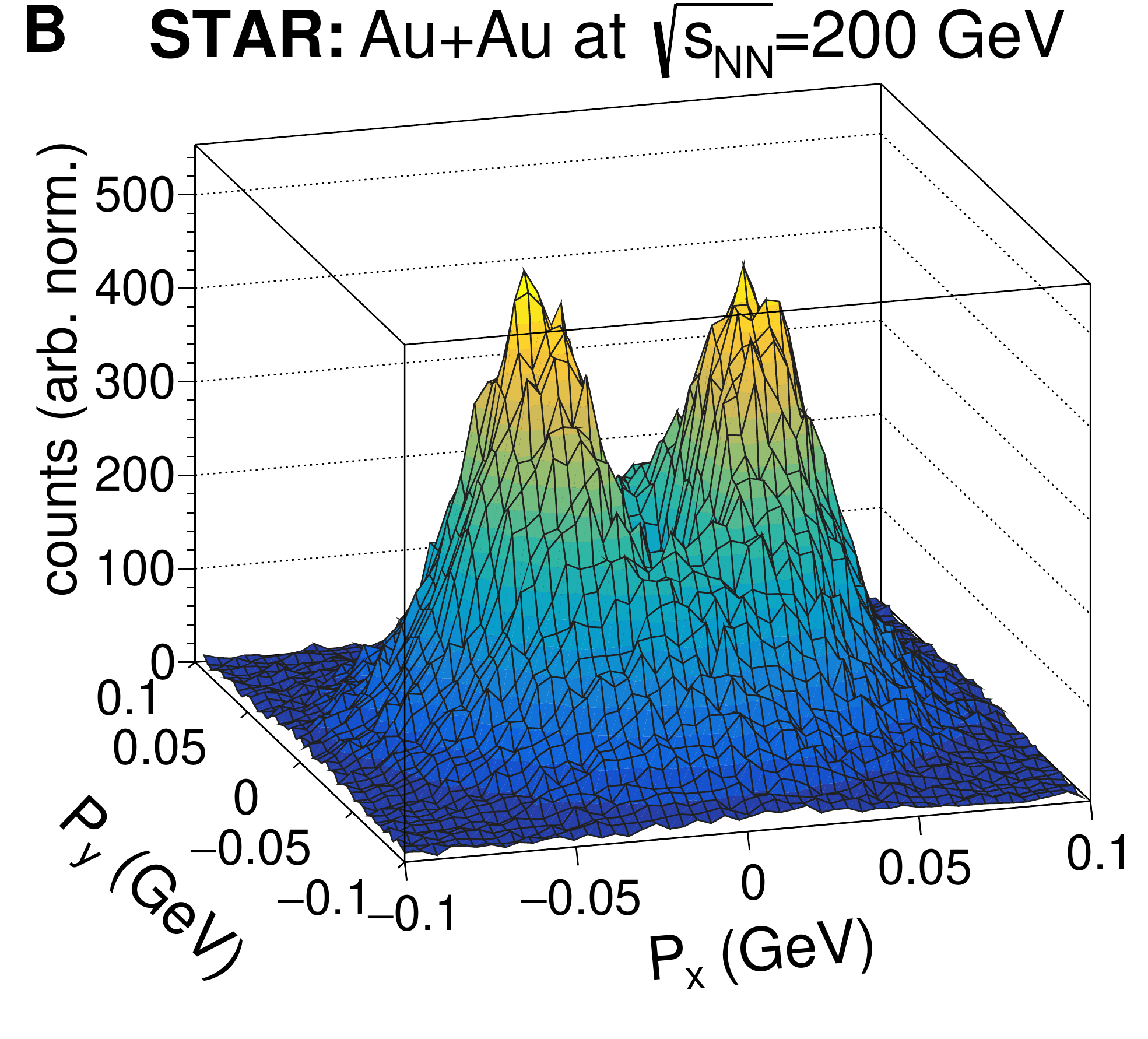}
    \includegraphics[width=0.49\textwidth]{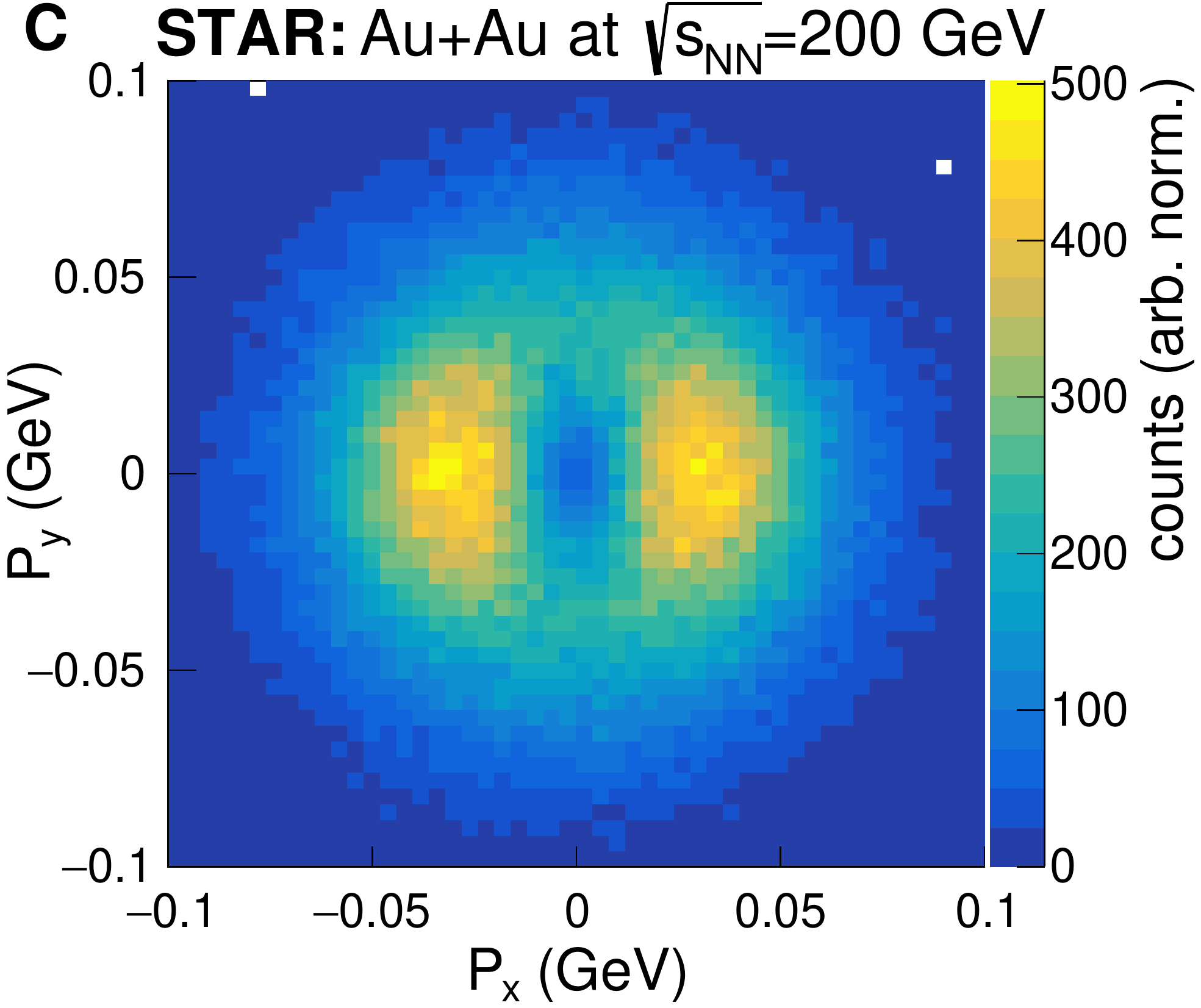}
    \includegraphics[width=0.49\textwidth]{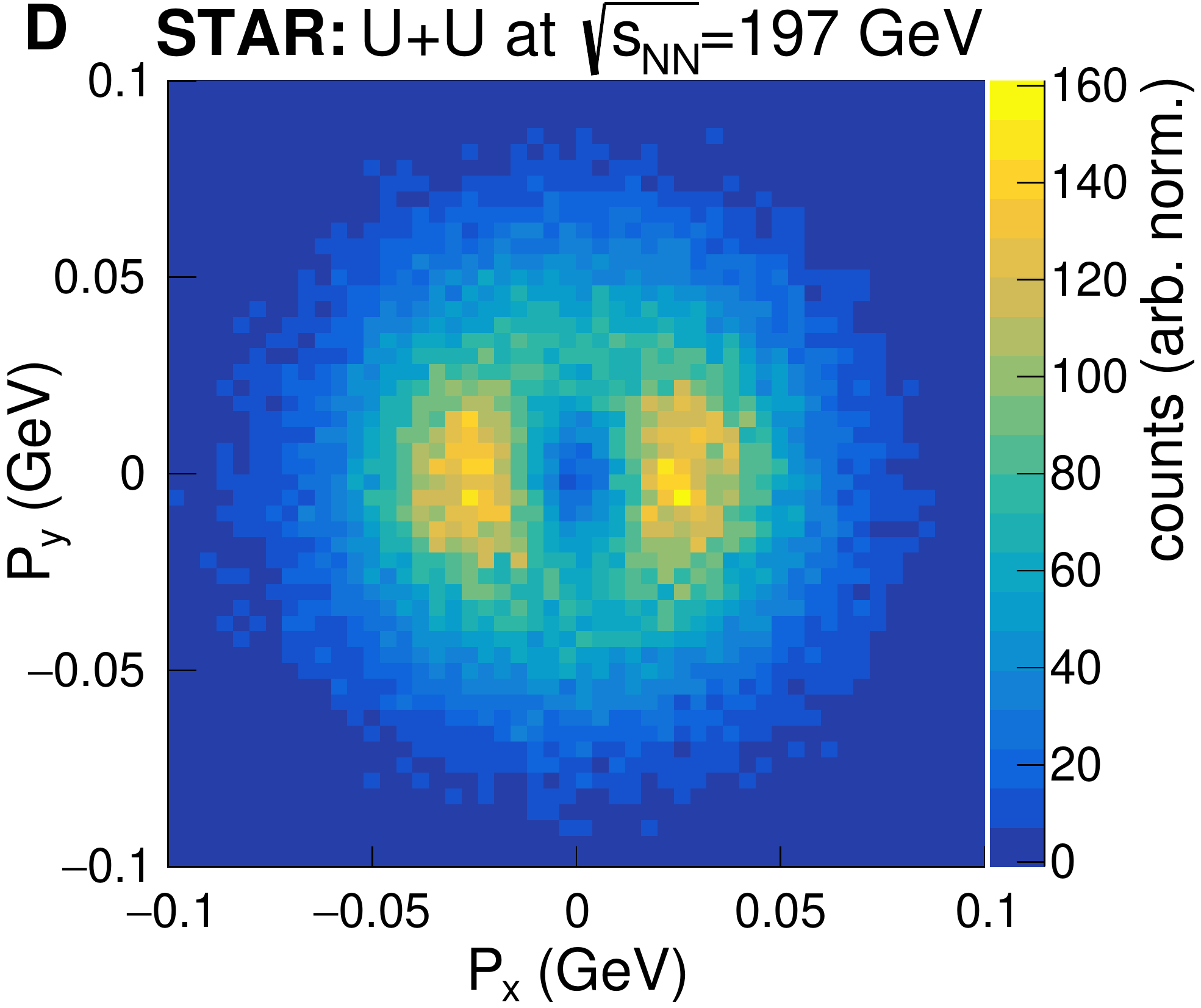}

    \caption{ \label{fig:fig_2} (A) The invariant mass distribution of $\pi^+\pi^-$ pairs collected from Au+Au and U+U collisions. The vertical black lines indicate the selected mass range with uniform detector acceptance and efficiency in $\phi$ that is used for all the subsequent analyses. Panels (B) to (D) show the two-dimensional distribution of the $\rho^0$ transverse momentum, where $P_x = P_{T}\cos\phi$ and $P_y = P_{T}\sin\phi$. The data from Au+Au collisions is shown in (B) as a continuous surface and in (C) as a 2D image. For comparison, the data from U+U collisions is also shown in (D) as a 2D image. In both Au+Au and U+U collisions, a clear asymmetry is visible due to the preferential alignment of the transverse momentum in the $x$ direction.
     }
     
    
\end{figure}

In this Article, the transverse linear polarization of photons in UPCs is used as an interferometry tool to explore the structure of heavy nuclei. The Article is structured as follows: First we discuss the selection of signal $\pi^+\pi^-$ pairs and the techniques used to reject backgrounds. Next we present the mathematical definition of the $\rho^0$ polarization observable, $\phi$, followed by a 2D investigation of the momentum of the $\rho^0$, by taking projections parallel and perpendicular to the polarization. Following a similar concept, we perform a 2D analysis of the $\rho^0$'s transverse momentum squared ($|t|$ distribution) as a function of the polarization angle, to demonstrate its profound effect on the apparent mass radius of the colliding nuclei. To gain an understanding of the nature of the interference effect, we then present a comparison of measurements in $p$+Au and Au+Au collisions at a center of mass energy per nucleon pair ($\sqrt{s_{_{NN}}}$) of 200 GeV and U+U collisions at $\sqrt{s_{_{NN}}} =$ 193 GeV. Next we present a method of taking this novel interference effect into account, thereby allowing the extraction of the 'true' mass radius, $R_0$, of the colliding nuclei. We end with a comparison and discussion of various theoretical models and potential explanations of the observed quantum interference measured through two non-identical particles.


\section*{RESULTS}
Figure \ref{fig:fig_2}A shows the measured invariant mass distribution for selected $\pi^+\pi^-$ pairs with $P_T < 200$~MeV from Au$+$Au and U$+$U collisions (note, we use capital $P$ to denote the momentum of the $\pi^+\pi^-$ pair, i.e. the $\rho^0$).
The data from Au$+$Au and U$+$U collisions were collected with the STAR detector at RHIC in the years of 2010 and 2011 while the data from $p+$Au collisions were taken in 2015 with additional tagging of the forward diffractive protons~\cite{STARRP:2020dzd}.
A broad prominent peak is observed in the invariant mass distribution around the $\rho^0$ mass of $\sim770$ MeV indicating that the trigger system and track selection criteria are effective for selecting exclusive $\rho^0\rightarrow \pi^+\pi^-$ processes. 
Candidate $\pi^+\pi^-$ pairs with an invariant mass ($M_{\pi\pi}$) between 650 and 900 MeV are selected for further analysis. Selecting this mass range results in a sample of $\pi^+\pi^-$ pairs predominantly from the decay of a $\rho^0$ vector meson produced in the diffractive photo-nuclear interaction.
The measured $\pi^+\pi^-$ final state also contains interfering contributions from other production channels with the same quantum numbers (e.g. continuum $\pi^+\pi^-$ from the Drell-S\"oding process~\cite{soding1966702} and $\rho^0-\omega$ mixing~\cite{TingPhysRevLett.27.888}). The relative amplitude contribution from each channel has previously been studied in detail by the STAR collaboration~\cite{star_collaboration_coherent_2017} and others~\cite{theh1collaborationDiffractiveElectroproductionRho2010,alicecollaborationCoherentPhotoproductionRho2020a}. In this analysis, the selected invariant mass range is used primarily to obtain a sample of $\pi^+\pi^-$ pairs with a large signal-to-background ratio and with roughly uniform acceptance. Throughout the article we follow the convention of referring collectively to the inseparable $\pi^+\pi^-$ final states with the same quantum numbers as $\rho^0$ candidates.  

\subsubsection*{Background and Contamination}
In addition to the $\rho^0$ which we study here, there are several other processes which potentially contaminate the selected sample.
The first potential contamination results from misidentified $\phi \rightarrow K^0_LK^0_S$ and the subsequent $K^0_S\rightarrow\pi^+\pi^-$ decay. Such background from $K^0_S$ is removed by requiring $M_{\pi\pi} > 650$ MeV.
Similarly, the $\phi$ meson and $\omega$ meson can produce $\pi^+\pi^-\pi^0$ final states, which contaminate the $\pi^+\pi^-$ sample since STAR does not detect $\pi^0$. 
Additional contamination may result from $\rho^\prime$ (e.g. $\rho(1450)$ or $\rho(1570)$) since it can decay to 4-pion final states ($\pi^+\pi^-\pi^+\pi^-$ and $\pi^+\pi^-\pi^0\pi^0$)~\cite{STARrhop:2009giy}. Since STAR does not measure $\pi^0$, any final state with two neutral pions appears as a $\pi^+\pi^-$ event. Similarly, any final state with four charged pions may contaminate the $\pi^+\pi^-$ signal pairs when two of the charged pions are lost, either due to imperfect reconstruction efficiency or by traveling outside of the acceptance.
All of these background processes are reconstructed predominately at low $M_{\pi\pi}$ and are therefore easily removed by analyzing $\pi^+\pi^-$ with $M_{\pi\pi} > 650$ MeV. For instance, analyses conducted at HERA~\cite{theh1collaborationDiffractiveElectroproductionRho2010} and Jefferson Laboratory (JLab)~\cite{CLAS:2008rpm,CLASELFASSI2012326} rejected these backgrounds by selecting candidates with $M_{\pi\pi} > 600$ MeV.
Despite the relatively long lifetime of the charged pion ($c\tau=7.8$ m) in comparison with the radius of STAR's primary tracking detector (R$\approx 2$ m), as many as 5\% of all the $\pi^+\pi^-$ pairs will have at least one of the pions undergo decay to a muon and a neutrino within that volume. These events, with a final state of $\pi^+\mu^-X$, $\mu^+\pi^-X$, or $\mu^+\mu^-X$ (where $X$ denotes the unmeasured (anti-)neutrino), originate from the same primordial process (e.g. diffractive photoproduction of a $\rho^0$ or direct $\pi^+\pi^-$ pair) that we study. However, they are undesirable for the measurement in question since the decay randomizes the angle and momentum of the measured final-state charged track.
The use of precise particle time-of-flight (TOF) measurement (see Materials and Methods) helps reject pairs in which one (or both) of the charged pions decay, especially for pairs where either daughter track has low transverse momentum ($p_T<300$ MeV). 
The lower invariant mass threshold ($M_{\pi\pi} > 0.65$ GeV) is useful for further reducing the contamination from such events since pairs with a daughter $\pi^\pm$ that decays to a $\mu^\pm$ and an (anti-)neutrino are generally reconstructed with an invariant mass shifted to a smaller value~\cite{alicecollaborationCoherentPhotoproductionRho2020a}.

\subsubsection*{Data Analysis}
An angular distribution, sensitive to photon polarization interference effects, is constructed from selected $\pi^+\pi^-$ pairs using the $\phi$ observable~\cite{Xing:2020hwh}, defined as:
\begin{linenomath}
\begin{equation}
    \cos{\phi} = (\vec{p_{T1}} + \vec{p_{T2}})\cdot{(\vec{p_{T1}} - \vec{p_{T2}})}/(|\vec{p_{T1}} + \vec{p_{T2}}|\times|\vec{p_{T1}} - \vec{p_{T2}}|)
\label{eq:phiangle}
\end{equation}
\end{linenomath}
where $\vec{p_{T1}}$ and $\vec{p_{T2}}$ are the 2D momentum vectors of the daughter pions in the plane transverse to the beam direction. At the relevant kinematics to this study with $|\vec{p_{T1}} + \vec{p_{T2}}|<<|\vec{p_{T1}} - \vec{p_{T2}}|$, $\phi$ angle from Eq.~\ref{eq:phiangle} is equivalent to the angle between the momentum of the parent momentum and the momentum of one of its daughters. Therefore we use these descriptions interchangeably throughout this Article. 
In order to remove any effect due to charge-dependent track reconstruction efficiency,
$\vec{p_{T1}}$ and $\vec{p_{T2}}$ are randomly assigned from the daughter $\pi^+$ and $\pi^-$ in each event. 
This has the additional effect of naturally eliminating any odd harmonics of the distribution. 
However, recent calculations have suggested that odd harmonics of the $\phi$ distribution may be sensitive to Coulomb-nuclear interference effects~\cite{hagiwaraCoulombNuclearInterference2021}, and will therefore be pursued in detail in future work. The transverse momentum distribution of the parent $\rho^0$ can be decomposed into two components, one parallel ($P_x$) to and one perpendicular ($P_y$) to the polarization direction: 
\begin{linenomath}
\begin{align}
    \label{eq:pxpy}
    \begin{split}
        P_x = P_T \cos{\phi}, \\
        P_y = P_T \sin{\phi}.
    \end{split}
\end{align}
\end{linenomath}
The result of this two-dimensional representation of the $\rho^0$ transverse momentum is shown in Figs.~\ref{fig:fig_2}B and \ref{fig:fig_2}C for Au+Au collisions and in Fig.~\ref{fig:fig_2}D for U+U collisions. 
In all three cases, a striking interference pattern is observed in the $P_x$ direction while no such effect is observed in the $P_y$ direction. As previously mentioned, the observation of an interference pattern along $P_x$ is expected, since the photon polarization is almost perfectly aligned with the impact parameter as shown in Fig.~\ref{fig:fig_1}C\cite{ZhaPhysRevD.103.033007}.

\begin{figure}
    \centering
    \includegraphics[width=0.49\textwidth]{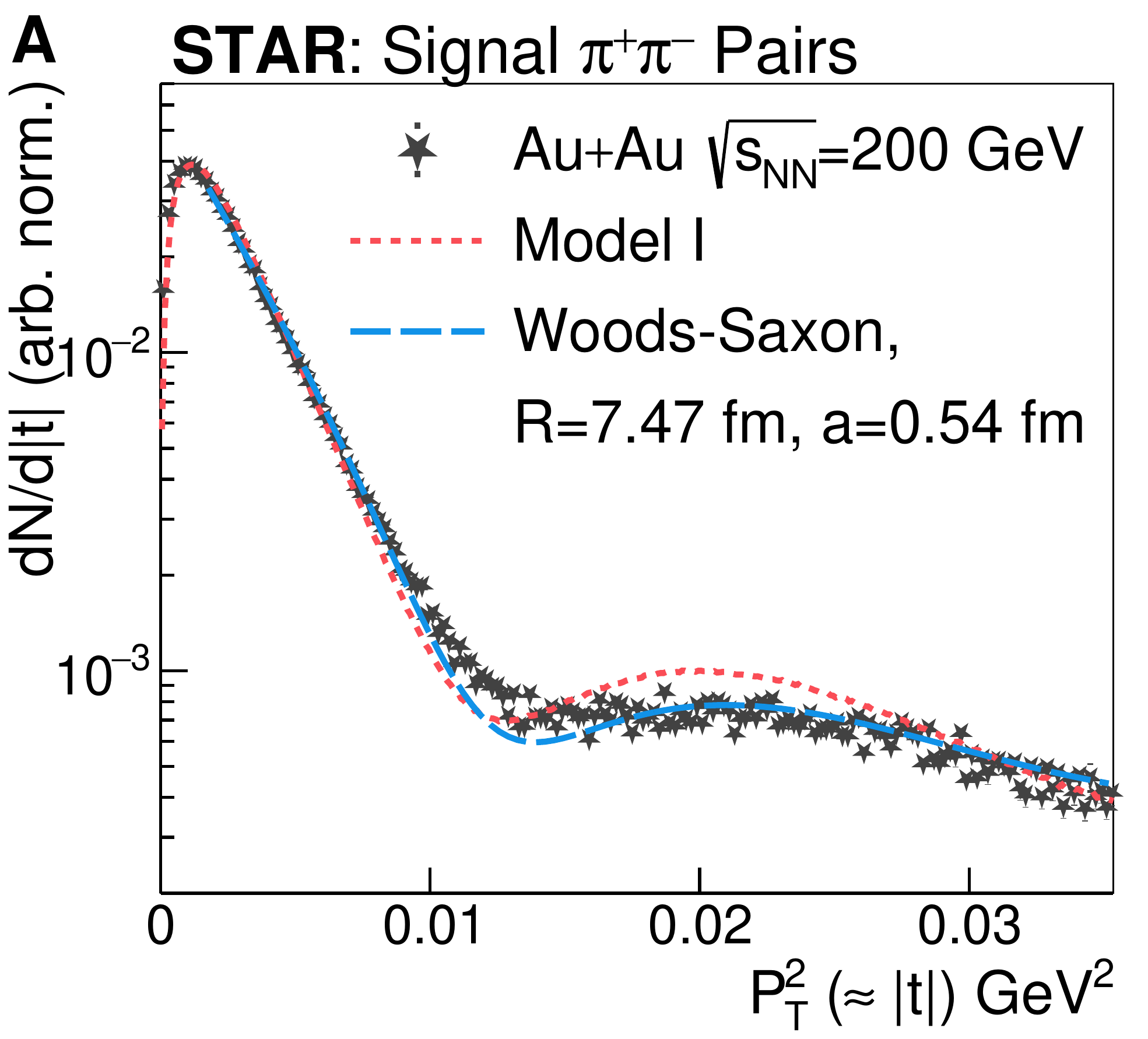}
    \includegraphics[width=0.49\textwidth]{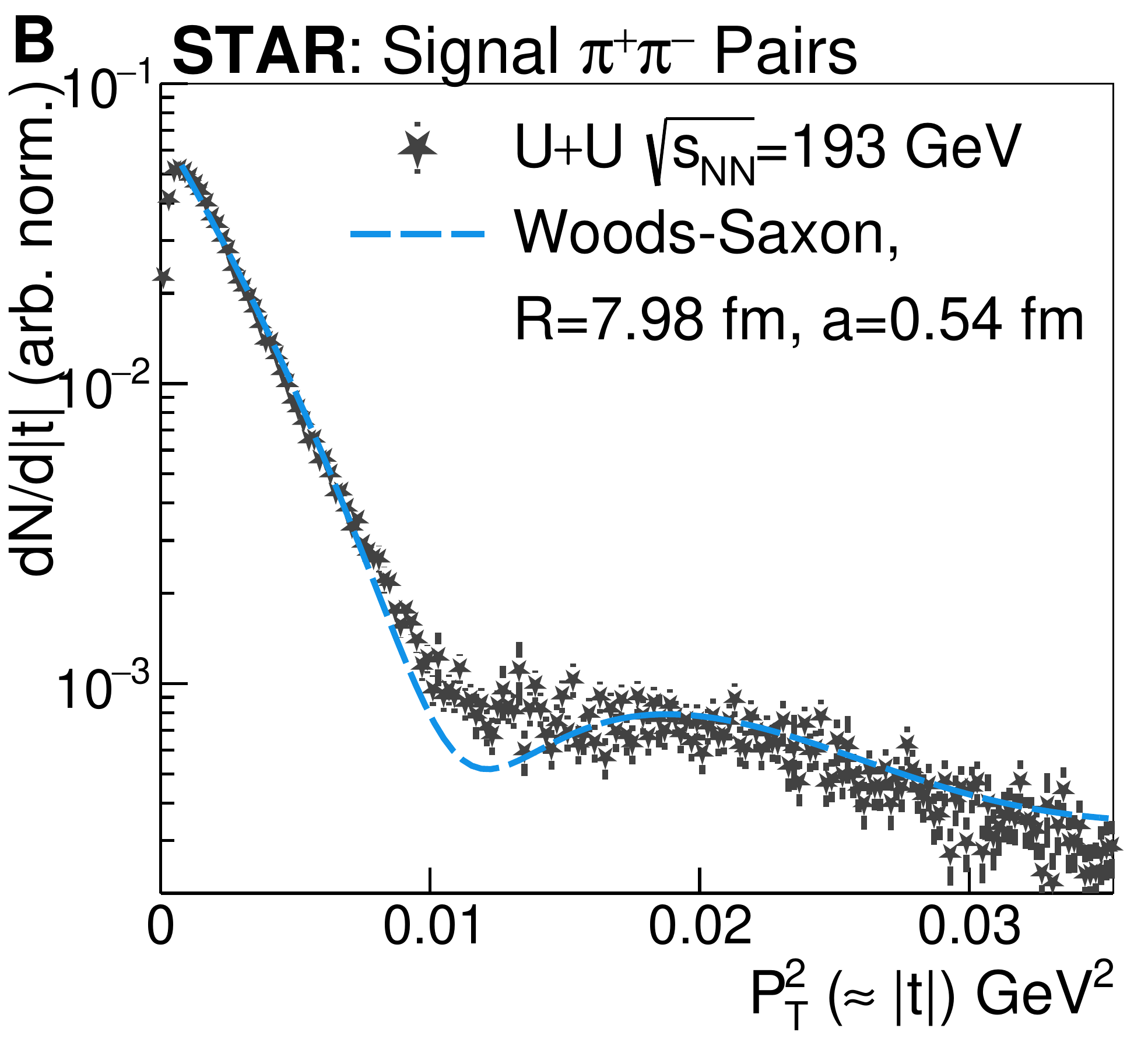}
    \includegraphics[width=0.49\textwidth]{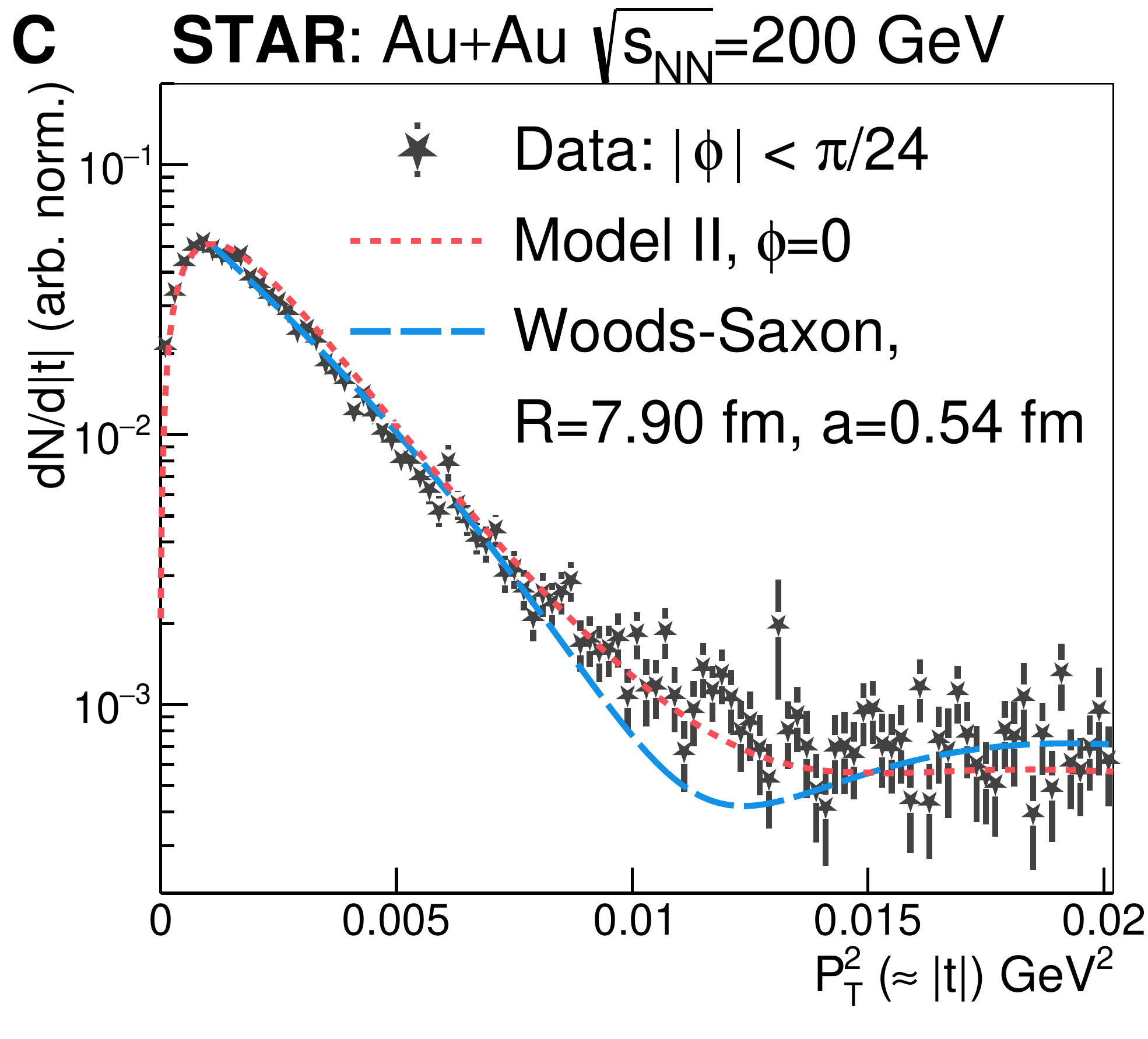}
    \includegraphics[width=0.49\textwidth]{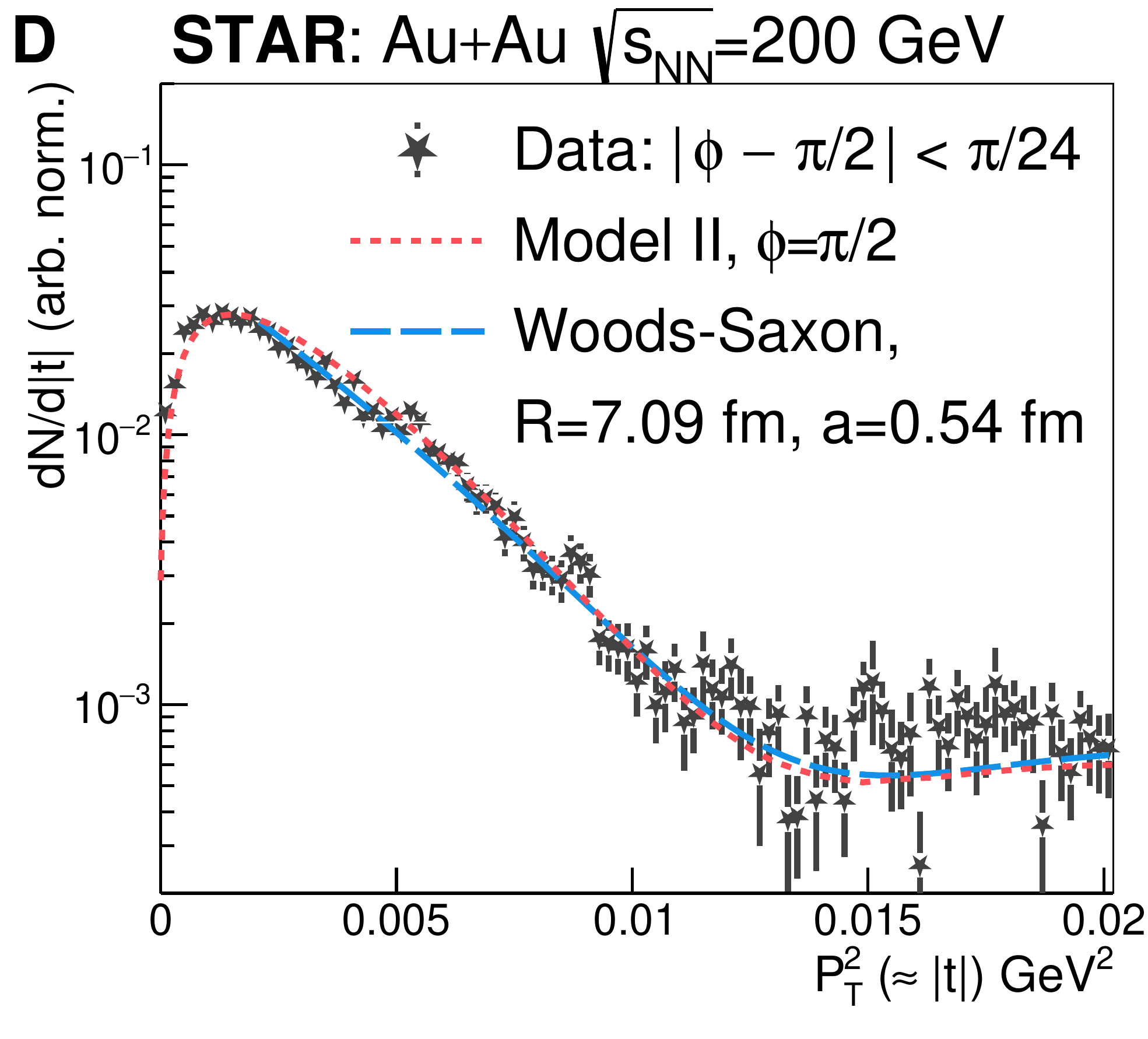}
    
    \caption{ \label{fig:fig_3} The $dN/d|t|$ as a function of $|t|$ ($\approx P_T^2$) for Au+Au collisions (A) and U+U collisions (B). The $dN/d|t|$ as a function of $|t|$ from Au+Au collisions for $\phi$ bins at $0^o$ (C) and $90^{\circ}$ (D). In (A-D) long-dashed curves are shown for the best-fit Woods-Saxon distribution. The distributions from Au+Au collisions (A, C, D) also show red short-dashed curves from two models. Model I~\cite{Zha:2018jin,ZhaPhysRevD.103.033007} is photon and Pomeron interaction with Woods-Saxon distributions using the $\rho^0N$ cross sections. Model II~\cite{Xing:2020hwh} is a dipole and gluon interaction with the gluon distribution inside nucleus in a color-glass condensate model including the $\rho^0$ wave function contribution. The distributions in (A-D) are not corrected for detector resolution or efficiency, nor are they normalized as absolute cross sections. The corrections were found to have a negligible effect on the shape of the distributions. Vertical error bars represent statistical uncertainties.  }

\end{figure}

Figures~\ref{fig:fig_3}A and 3B show the measured $P_T^2\approx|t|$ distribution for signal $\pi^+\pi^-$ pairs from Au+Au (A) and U+U (B) collisions with two distinct classic diffractive peaks. A clear difference between the shape of the two spectra is visible, especially for $|t| < 0.01$ (GeV)$^{2}$ due to the different size and shape of the two nuclei. 
The measured $|t|$ spectra can be described by an empirical model, composed of a coherent contribution characterized by the form factor of a Woods-Saxon distribution and an incoherent contribution characterized by a dipole form factor.  The Woods-Saxon distribution, $\rho_A(r)$ is commonly used to describe the density distribution of spherically symmetric heavy nuclei. It has the form:
\begin{equation}
    \label{eq:ws}
    \rho_A(r; R, a) = \frac{ \rho_0 }{ 1 + \exp[ (r - R)/a ] },
\end{equation}
where $R$ is the nuclear radius, $a$ is the surface thickness, and the factor $\rho_0$ is a normalization factor defined as $\rho_0 = 3A/(4\pi R^3)$, with $A$ being the atomic mass number for the given nucleus.
The nuclear form factor can be computed from the density distribution via the Fourier Transform, denoted in Eq.~\ref{eq:fit_ws} below using the notation $\mathcal{F}[...](k)$ (see Materials and Methods for more details).
Following this approach, the best-fit Woods-Saxon radius for both gold and uranium can be extracted from the $|t|$ spectra by fitting them to a function of the form:
\begin{linenomath}
\begin{equation}
    \label{eq:fit_ws}
    f(t) = A_{c}\lvert \mathcal{F}\left[ \rho_A(r; R, a) \right](|t|) \rvert^{2} + \frac{A_{i}/Q_0^2}{ (1 + |t|/Q_0^2)^2 },
\end{equation}
\end{linenomath}
where $A_c$ is the amount of coherent production and $A_{i}$ is the amount of incoherent production characterized by a dipole form factor. The value of $Q_0^2$ is fixed to $Q_0^2=0.099$ ${\rm GeV}^{-2}$ from previous measurement of the incoherent contribution~\cite{star_collaboration_coherent_2017, Schafer:2020bnm}. 
Figures~\ref{fig:fig_3}C and D further show the $|t|$ distribution for $\lvert\phi\rvert < \pi/24$ (C) and $\lvert\phi-\pi/2\rvert < \pi/24$ (D) to illustrate the profound impact that the interference effect has on the $|t|$ distribution. Indeed, the extracted radius for the case with maximum interference ($\phi\simeq$0) is 
larger by $0.81\pm0.05({\rm stat.}) \pm0.03 ({\rm syst.})$ fm compared to the case with minimum interference ($\phi\simeq\pi/2$).

\begin{figure}
    \centering    
    \includegraphics[width=0.48\textwidth]{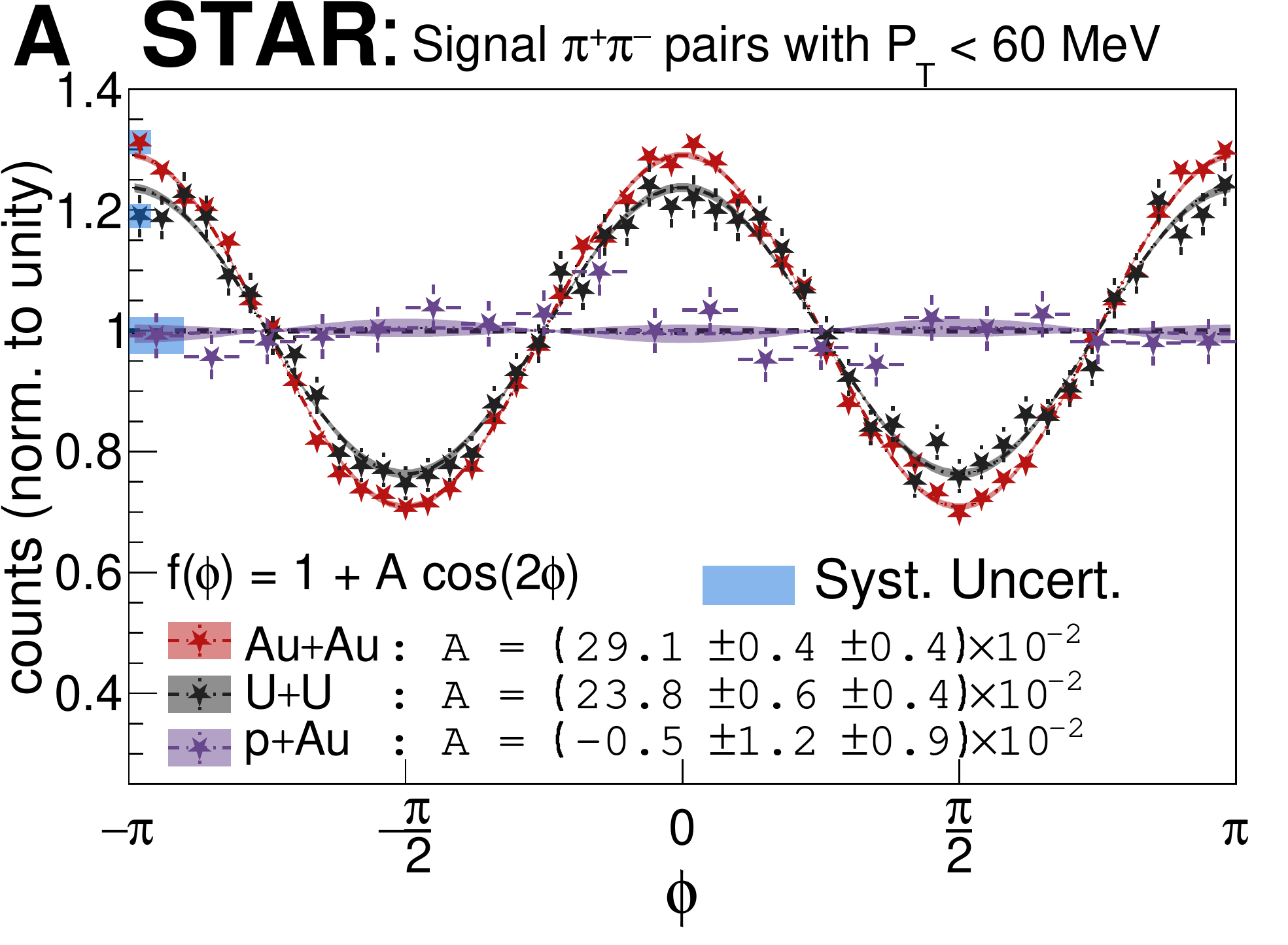}
    \includegraphics[width=0.48\textwidth]{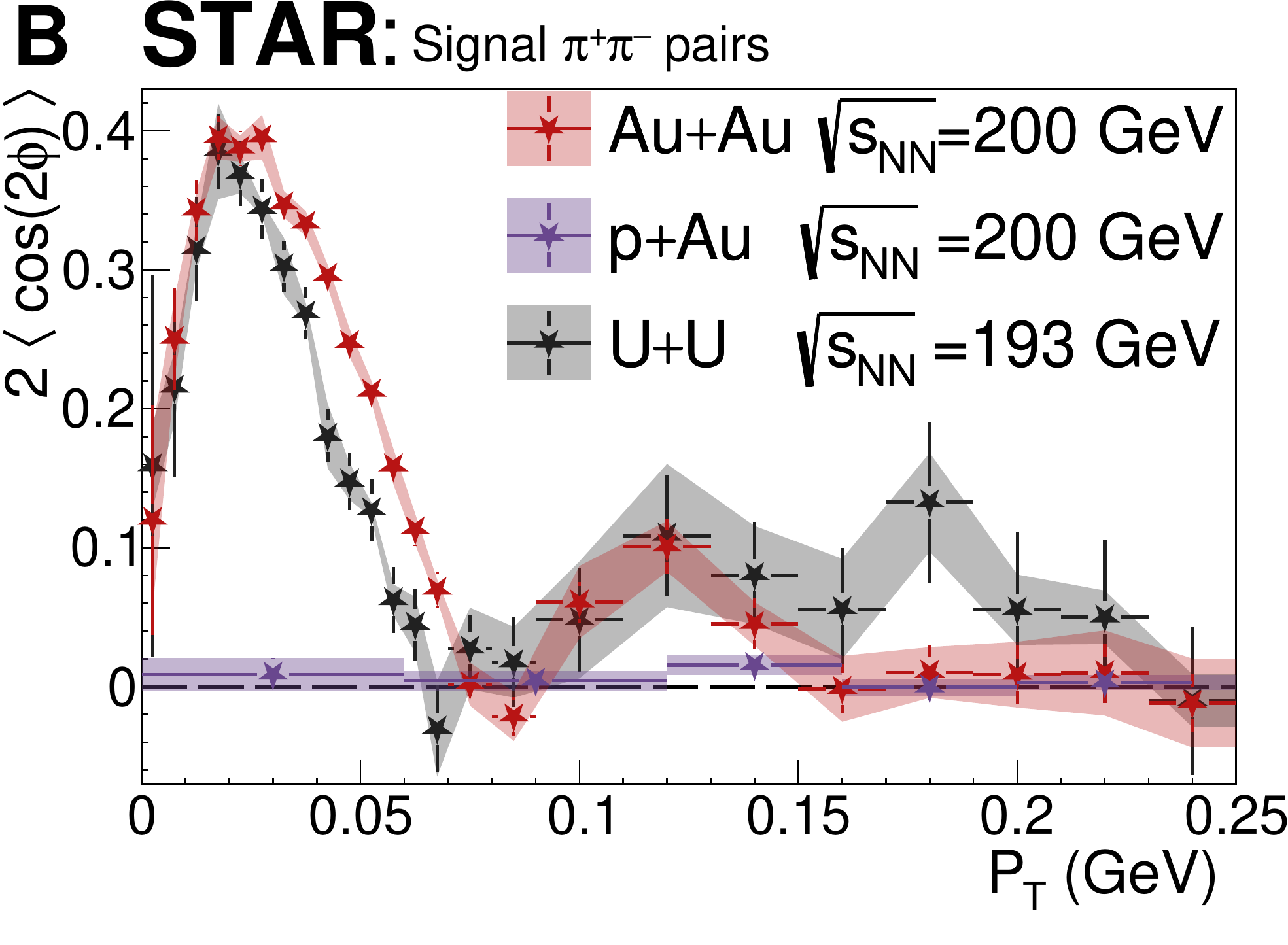}
    
    \caption{ \label{fig:fig_4} (A) The $\phi$ distribution for $\pi^+\pi^-$ pairs collected from Au+Au and U+U collisions with a pair transverse momentum ($P_T$) less than 60 MeV and an invariant mass between 650 and 900 MeV. Statistical uncertainties are shown as vertical bars on all points, while systematic uncertainties are shown as filled boxes only on the leftmost points to improve the clarity.  The $\phi$ distributions are fit to a function of the form $f(\phi) = 1 + A \cos2\phi$ to extract the amplitude ($A$) of the $\cos2\phi$ modulation. The quoted uncertainties on $A$ are for statistical and systematic sources of uncertainty, respectively. (B) The fully corrected $2\langle \cos2\phi\rangle$ modulation vs. $P_T$ for Au+Au and U+U collisions. The statistical uncertainty on each data point is shown in vertical bars with the systematic uncertainty shown in the shaded bands. }
\end{figure}

One can also examine the $\phi$ distribution itself. 
The $\phi$ distribution for Au+Au, U+U, and p+Au events is shown in Fig.~\ref{fig:fig_4}A for pairs with $650 < M_{\pi\pi} < 900$ MeV and with $P_T < 60$ MeV. 
Each distribution is scaled such that the average yield (integrated over $\phi$) is unity, to improve comparison. 
The low $P_T$ range is chosen to select a region where coherent production is dominant (in the case of Au+Au and U+U), with very little production from incoherent photo-nuclear interactions. 
The signal in this low $P_T$ range is almost completely unaffected by the STAR detector acceptance and efficiency. 
In both the Au+Au and U+U data sets, a clear and prominent $\cos2\phi$ modulation is visible while the p+Au data displays an isotropic distribution. 
The amplitude of the modulation is quantitatively determined by fitting the distribution to a function of the form:
\begin{linenomath}
\begin{equation}
    f(\phi) = 1 + A \cos 2\phi
\end{equation}
\end{linenomath}
where $A$ is the amplitude of the $\cos2\phi$ modulation. 
The extracted amplitudes are \\$A_\mathrm{Au+Au}~=[29.2~\pm0.4~\mathrm{(stat.)}~\pm0.4~\mathrm{(syst.)}]\times10^{-2}$ ($\chi^2/\mathrm{ndf}=45/49$), $A_\mathrm{U+U}~=[23.7~\pm0.6~\mathrm{(stat.)}~\pm0.4~\mathrm{(syst.)}]\times10^{-2}$ ($\chi^2/\mathrm{ndf}=34/49$), and $A_\mathrm{p+Au}~=[-0.5~\pm1.2~\mathrm{(stat.)}~\pm0.9~\mathrm{(syst.)}]\times10^{-2}$ ($\chi^2/\mathrm{ndf}=16/19$) for Au+Au, U+U, and p+Au collisions, respectively. The fully corrected $2\langle \cos 2\phi \rangle$ distributions as a function of the $\rho^0$ transverse momentum are shown in Fig.~\ref{fig:fig_4}B for Au+Au, U+U, and p+Au collisions. 
In both the Au+Au and U+U cases, a prominent maximum is visible at low $P_T$ which falls rapidly towards zero with another smaller peak visible at higher $P_T$. 
In contrast, the data from p+Au collisions shows no structure, being consistent with zero at all $P_T$.
At the highest reported $P_T$ ($\sim 240$ MeV) the modulation strength is consistent with zero in all three data sets. 

To study the impact of photon momentum and interference, we perform another analysis by dividing the $|t|$ distribution into different $\phi$ bins. In each $\phi$ bin, the $|t|$ distribution is fit with Eq.~\ref{eq:fit_ws}. Examples of two $\phi$ slices ($\phi=0^\circ,90^\circ$) of the cross section as a function of $|t|$ are shown in Figs.~\ref{fig:fig_3}C and \ref{fig:fig_3}D, together with the empirical fit (using Eq.~\ref{eq:fit_ws}) and predictions from Model II. In principle, Eq.~\ref{eq:fit_ws} should be able to describe the entire $|t|$ spectra, except the very low-$|t|$ region where detector resolution, photon transverse momentum and destructive interference effects are dominant. Indeed, calculations shown as Model I and II in Fig.~\ref{fig:fig_3} which take these effects into account are able to describe the entire spectrum, including the low $|t|$ dip quite well. 
Figure~\ref{fig:fig_5}A presents the radius $R$ extracted by fitting Eq.~\ref{eq:fit_ws} to data as a function of $\phi$. There appears to be a second-order modulation in the resulting $R$ as a function of $\phi$. Therefore the equation 
\begin{linenomath}
\begin{equation} 
    \label{eq:Rcos2phi}
    R = \sqrt{R_0^2 + \sigma_b^2/\epsilon_p\times( 1 + \epsilon_p\cos{2\phi})}
\end{equation}
\end{linenomath}
is used to extract the minimum radius $R_0$ at $\phi=90^\circ$ with a second-order angular modulation parameter $\sigma_b$ where the parameter $\epsilon_p=\langle\cos{2\phi}\rangle=1/2$ is included to account for the resolution of the measurement when using the daughter momentum as a proxy for the polarization direction.
The $\sigma_b$ parameter quantifies the observed strength of the interference effect, taking into account multiple contributions from effects such as: the range in impact parameter probed, the finite photon transverse momentum, and the resolution of the polarization measurement. 
The detailed contribution of each effect to the final value of $\sigma_b$ can be ascertained from Model I and Model II, but is out-of-scope for this Article. 
These details, their finer effects on the shape of the $|t|$ distribution, and the final strength of the modulation may be further explored in the future.
Using this approach which gives the value of the true diffractive radius from the nuclear form factor, we obtain $R_0 = 6.62\pm 0.03$ fm, $a = 0.5 \pm 0.1$ fm, and $\sigma_b = 2.38 \pm 0.04$ fm for Au+Au, and $R_0 = 7.37\pm 0.06$ fm, $a = 0.5 \pm 0.1$ fm, and $\sigma_b = 1.9 \pm 0.1$ fm for U+U. Quoted uncertainties are those for Poisson statistics. Systematic uncertainties are computed for the fully corrected values discussed below (see Materials and Methods).
\section*{Discussions and Comparison to Theory}

This measurement reports on the observation of a prominent $\cos 2\phi$ modulation observed in the $\rho^0 \rightarrow \pi^+\pi^-$ photoproduction process measured in Au+Au and U+U collisions. The modulation is observed in both Au+Au and U+U collisions but not in p+Au collisions using the same techniques.  
Unlike A+A collisions, which can undergo photo-nuclear interactions via the amplitudes depicted by the two diagrams shown in Fig.~\ref{fig:fig_1}A, the drastic difference in charge ($Z$) between the proton ($Z=1$) vs. the Au nucleus ($Z=79$) dominantly occurs through the one with $\gamma^{\rm Au} + \mathbb{P}^{\rm p} \rightarrow \rho^0$, i.e. where the photon originates from the electromagnetic field of the gold nucleus, and the Pomeron originates from the proton. According to STARLight~\cite{klein_starlight_2017}, for pairs within STAR acceptance with $P_T < 100$ MeV, the case in which the photon comes from the field of the Au and the Pomeron is emitted by the proton (See Fig.~\ref{fig:fig_1}B) accounts for 91\% of the observed cross section. The other 9\% result from the other case, with $\gamma^{\rm p} + \mathbb{P}^{\rm Au} \rightarrow \rho^0$ (p+Au), where the photon comes from the electromagnetic field of the proton and the Pomeron from the Au nucleus. As mentioned, the main reason for this drastic difference in the amplitude of the two processes is due to the charge difference. There is also a dramatic difference in the $|t|$ distribution due to the form factor from the nuclear (proton) size. 
For these reasons, the p+Au collisions are essentially free from interference effects as depicted in Fig.~\ref{fig:fig_1}B.
On the other hand, for symmetric collisions of highly charged nuclei like Au or U, the amplitudes for the two diagrams shown in Fig.~\ref{fig:fig_1}A are equal and therefore lead to maximal interference.
Therefore, the observation of a prominent $\cos 2\phi$ modulation in Au+Au and U+U collisions, while absent in p+Au collisions, is consistent with the expectation, if quantum interference is the source of the modulation.

In the past, several $\rho^0$ polarization measurements have been carried out at HERA~\cite{theh1collaborationDiffractiveElectroproductionRho2010}, JLab~\cite{CLAS:2008rpm}, RHIC~\cite{starcollaborationCoherentEnsuremathRho2002}, and elsewhere via measurement of the spin density matrix in the helicity frame~\cite{schillingAnalysisVectormesonProduction1970,schillingHowAnalyseVectormeson1973a}. 
Past measurements, e.g. at HERA and JLab which measured an angular modulation with respect to the scattered lepton, involved highly virtual photons due to large momentum transfers that were therefore predominantly longitudinally polarized. 
Unlike those previous measurements, this $\cos 2\phi$ modulation results specifically from interference and is observable even when the real photon momentum and polarization vector are unknown, since the quantum interference leads to a correlation between the daughter pions momenta and the pion pair ($\rho^0$) momentum. It should also be noted that at $P_T=0$ the helicity frame representation and the $\phi$ angle coincide, but are not identical at finite $P_T$. 

Since this $\cos 2\phi$ modulation is a consequence of interference between the two amplitudes depicted in Fig.~\ref{fig:fig_1}A, it should be sensitive to the details of the photon and Pomeron wavefunctions, and therefore, potentially sensitive to the nuclear geometry and gluon distribution inside the colliding nuclei.
As shown in Fig.~\ref{fig:fig_4}A, the amplitude of the modulation for $P_T < 60$~MeV is stronger by about $5.5\%$ in Au+Au collisions than in U+U collisions with a 4.3$\sigma$ significance, including statistical and fully uncorrelated systematic uncertainty. 
If the systematic uncertainty is considered to be 100\% correlated between the two measurements, then the significance is 7.6$\sigma$. 
This significant difference at low $P_T$ (where the cross-section is dominated by the coherent photoproduction process) between gold and uranium already demonstrates sensitivity to the geometry of the colliding nuclei.
However, the details of the interference effect are more clearly seen from the $P_T$ dependence of the $2\langle\cos 2\phi\rangle$ distribution shown in Fig.~\ref{fig:fig_4}B.
While the Au+Au and U+U curves have a maximum $\cos2\phi$ modulation strength at approximately the same value ($\approx40\%$), the Au+Au curve has a wider first peak, leading to the larger average strength observed for $P_T < 60$~MeV.
However, at higher $P_T$ the relative contribution from coherent photoproduction rapidly decreases as the contribution from incoherent photoproduction takes over. 
The strength of the $\cos2\phi$ modulation is consistent with zero for $P_T>200 $ MeV, where the measured distribution is dominated by incoherent production. 
In the range between these two regions, at intermediate $P_T$ ($P_T \approx 120$ MeV), a second peak is observed in the $\cos2\phi$ modulation strength in both Au+Au and U+U collisions. While the structure of the $\cos{2\phi}$ modulation vs. $P_T$ is similar to the diffractive structure $d\sigma/dP_T$, we note that the first peak of the $\cos{2\phi}$ modulation is narrower than the first peak in $d\sigma/dP_T$ and the second peak of the $\cos{2\phi}$ distribution is approximately where the first dip is in $d\sigma/dP_T$.
There is some indication that the second peak visible in the U+U data is broader than the second peak in the Au+Au data, possibly due to the non-spherical geometry of the uranium nucleus or possibly due to differences in neutron skins between the two nuclei. However, with the relatively large uncertainties at $P_T > 100$ MeV, no significant difference in the second peaks of U+U vs. Au+Au can be claimed at this time.    

The extracted values of $R_0$ from the $|t|$ distributions as a function of $\phi$ should have the least contribution from photon transverse momentum and interference effects. However, there are still contributions from the finite size $\rho^0$ wavefunction~\cite{Forshaw:2010py}, with $R_T^{\rho}=1.03$ fm, and the finite angle between the photon polarization and impact parameter illustrated in Fig.~\ref{fig:fig_1}C. The depolarization can be estimated from the average angle between the photon and the impact parameter as $1-P=(R_0/\langle b\rangle)^2/4\simeq0.03$. The resulting true nuclear radii are  $R_A^2=R_0^2-R^{\rho}_T{}^2-(1/P-1)\sigma_b^2$ resulting in $R_{\rm Au}=6.53\pm0.03$ (stat.) $\pm0.05$ (syst.) fm for $^{197}Au$ and $R_{\rm U}=7.29\pm0.06$ (stat.) $\pm0.05$ (syst.) fm for $^{238}U$. 
The systematic uncertainty is dominated by the difference resulting from the use of different form factors and fit ranges. 
This also shows that, unlike $\rho^0$ photoproduction off nucleons or light nuclei, the details of the $\rho^0$ wave function (size) actually plays almost no role in determining the nuclear radii. The ratio of these two nuclear radii is $1.12\pm0.01$. Furthermore, these radii are systematically larger than the nuclear charge radii obtained from low-energy electron scattering~\cite{Shou:2014eya}. 
It should be noted that the strong-interaction nuclear radii have been measured at other facilities at lower energies~\cite{Alvensleben:1970uw,Alvensleben:1970uv,CornellPhysRevD.4.2683} with diffractive photoproduction of $\rho^0$ in photon-nucleus collisions. A scaling of $R=(1.12\pm0.02)A^{1/3}$ was extracted from the experimental data at DESY~\cite{Alvensleben:1970uw,Alvensleben:1970uv} with $R_{\rm Au}=6.45\pm0.27$ ${\rm fm}$ and $R_{\rm U}=6.90\pm0.14$ ${\rm fm}$ while another experiment at Cornell~\cite{CornellPhysRevD.4.2683} obtained $R_{\rm Au}=6.74\pm0.06$ ${\rm fm}$. The neutron skin~\cite{Tsang:2012se}, a root-mean-square difference between the neutral and nuclear charge radii, has been measured in even-even stable nuclei and shown to be around 0.3 ${\rm fm}$ for $^{208}Pb$~\cite{PREXPbSkinPhysRevLett.126.172502}. When the strong-interaction radius is taken as a weighted average of neutral and nuclear charge radii, our measurements of the same quantity yield: $S = R_n - R_p = 0.17 \pm 0.03 {\rm(stat.)} \pm 0.08 {\rm(syst.)}$ fm for $^{197}Au$ and $0.44\pm0.05 {\rm(stat.)} \pm 0.08{\rm(syst.)}$ fm for $^{238}U$ (using previous measurements of $R_p$~\cite{devriesNuclearChargedensitydistributionParameters1987,Shou:2014eya}). Our measurement of $S$ for $^{197}Au$ seems to follow the trend of world measurements at low energies~\cite{CentellesPhysRevLett.102.122502} while that of $^{238}U$ is significantly non-zero and indicates a value larger than that expected for neutron skins of similar nuclei [in terms of the fraction of neutron excess (N-Z)/A] ~\cite{CentellesPhysRevLett.102.122502}.

\begin{figure}
    \centering    
    \includegraphics[width=0.48\textwidth]{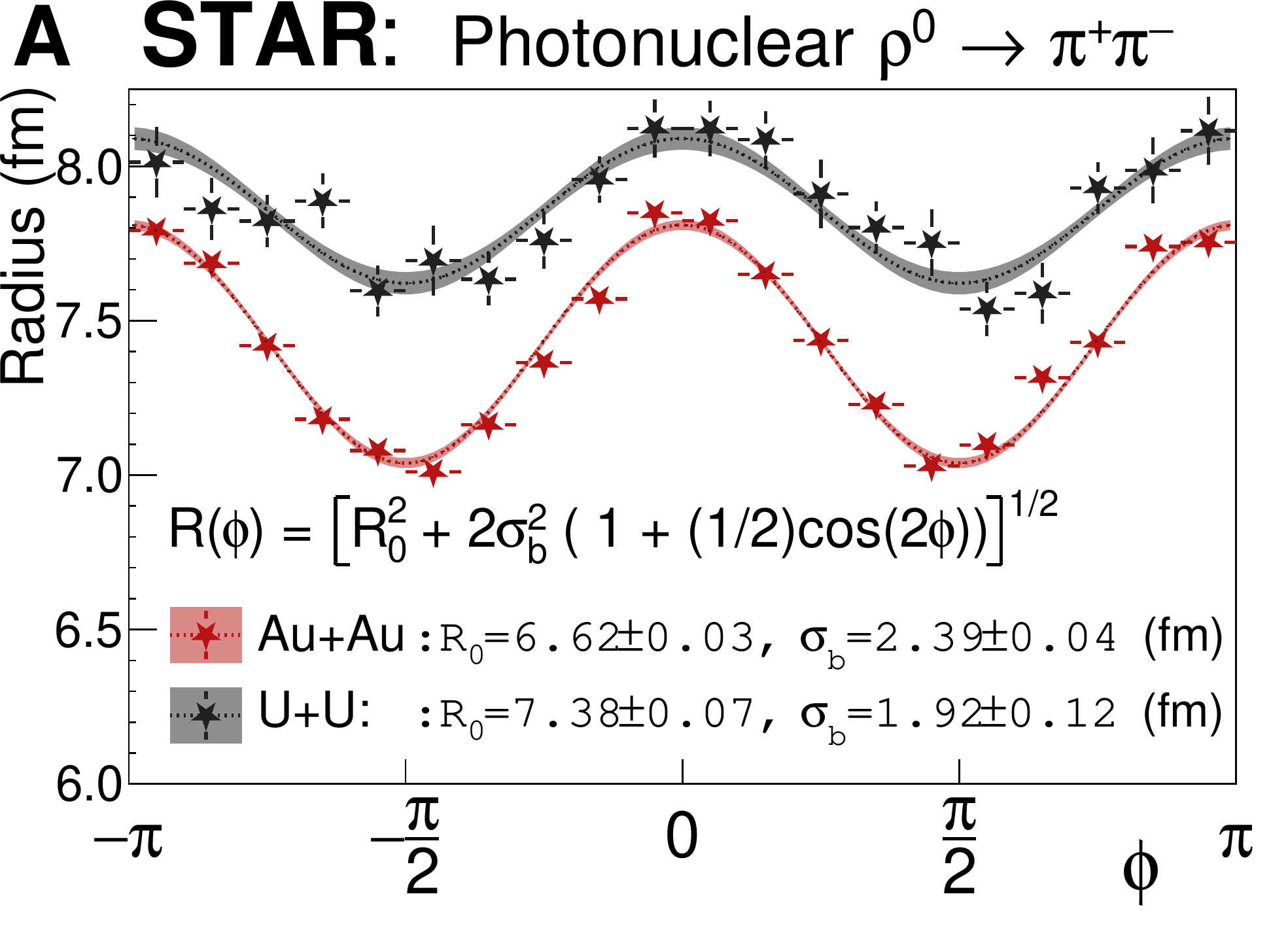}
    \includegraphics[width=0.48\textwidth]{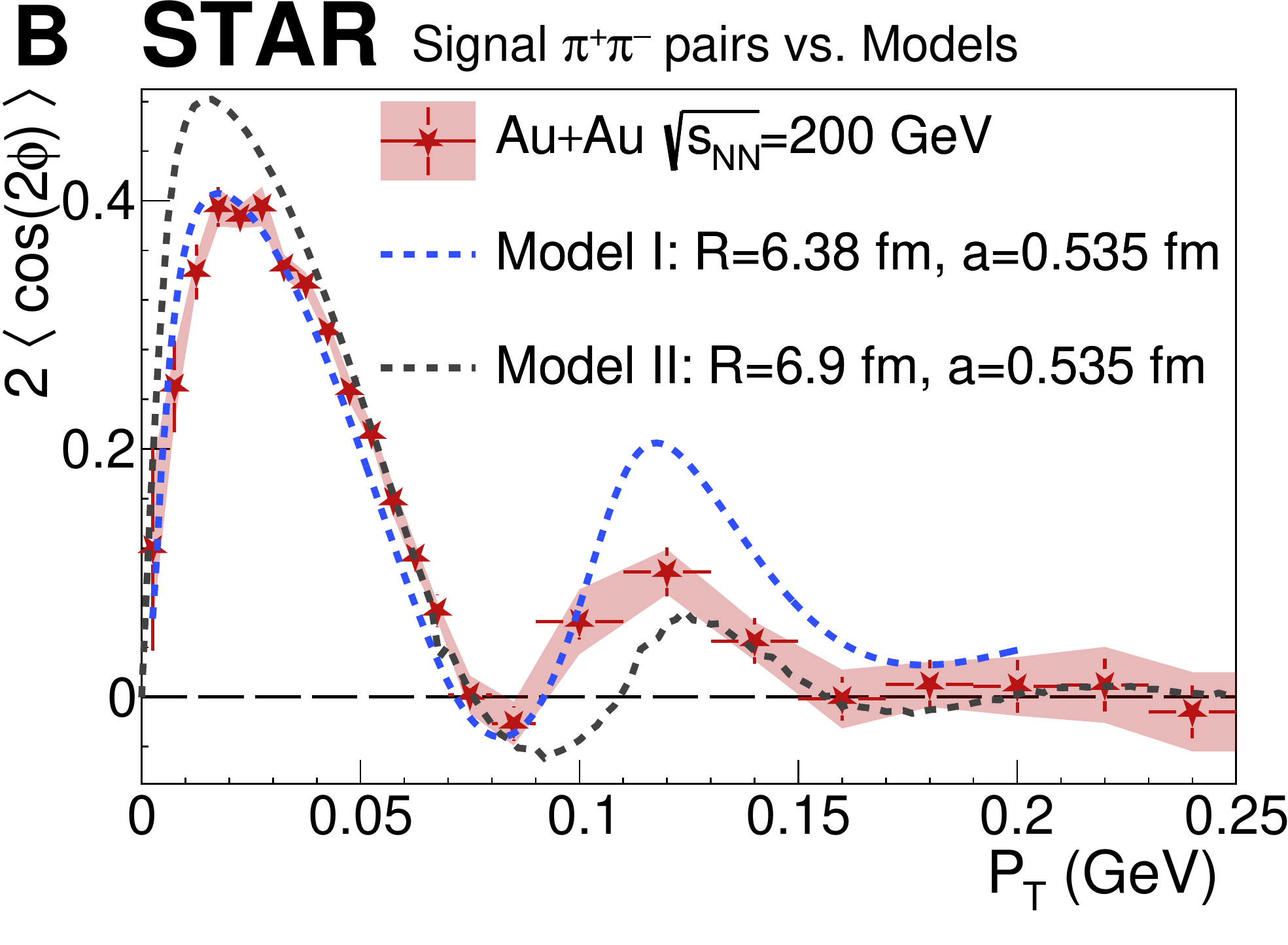}

    \caption{
    (A) Radial parameter as a function of the $\phi$ angle for Au+Au and U+U with an empirical second order modulation fit.  
    (B) Comparison between the fully corrected Au+Au distribution and theoretical calculations~\cite{Xing:2020hwh,ZhaPhysRevD.103.033007} that include the photon's linear polarization and two-source interference effects. 
    \label{fig:fig_5}
    }
\end{figure}

We also compare to theoretical calculations from Model I and Model II that take into account the interacting photon's transverse linear polarization and the quantum interference effects. 
Fig.~\ref{fig:fig_3} shows very good agreement between the data and both models over the entire $|t|$ distributions except for small disagreements in the structures at higher $|t|$.  
Figure~\ref{fig:fig_5}B shows a comparison of the $\cos{2\phi}$ modulation between the Au+Au data and two different theoretical calculations from Model I and II. 
Both of the model calculations are able to reproduce the qualitative features of the data, namely the prominent peak at low $P_T$ and the second peak at higher $P_T$. 
Model I predicts a peak modulation strength and position of the first peak in good agreement with the data while Model II does not reproduce the precisely measured magnitude or structure as observed. 
While neither model describes the precise location and amplitude of the second peak, both theoretical models indicate that the detailed structure of the modulation is sensitive to the distribution of gluons within the nucleus. In fact, Model II used gluon saturation and polarization with an effective nuclear radius of $6.9$ fm to reproduce the data shown in Fig.~\ref{fig:fig_3} and  Fig.~\ref{fig:fig_5}B. Although CGC is an effective theory applicable to the non-perturbative regime of QCD, it usually requires an energy scale above $\sim$1 GeV for reasonable leading-order calculation as depicted in Ref~\cite{Xing:2020hwh} and in Fig.~\ref{fig:fig_1}A. 
Considering that the $\rho^0$ mass is $\sim770$ MeV and close to that energy scale, it may not provide the hard scale needed for reliable leading-order model calculation. Additional phenomenological approaches have been demonstrated in the modelling of color transparency~\cite{JAINColor199667,CLASELFASSI2012326} and may be needed in this case to achieve the necessary precision for quantitative comparison to data.
The full 2D $|t|$ distribution as a function of $\phi$ explored in this study contains rich information about the nuclear geometry and gluon distribution at small $x$, which could be further investigated at RHIC, LHC, and EIC. 

As illustrated in Fig.~\ref{fig:fig_1}A, the observability of the $\rho^0$ spin alignment is a result of the interference of two wave functions from two indistinguishable target and projectile nuclei at the distance of an impact parameter apart from each other. This scheme is very different from most of the fundamental particle interactions where two wave functions are connected by a mediator or virtual particle in the corresponding Feynman diagram~\cite{KleinPT:2017}. In this case, the two potential $\rho^0$ wave functions only overlap through the decay daughters that propagate to the detector. There are several possible scenarios that would result in interference effects. First, if the phases of these $\rho^0$ are random, this would be similar to the azimuthal Hanbury Brown and Twiss intensity interference effect~\cite{magana-loaizaHanburyBrownTwiss,Baym:1997ce,Lisa:2005dd}. In this circumstance, both coherent and incoherent $\rho^0$ production would create a correlation~\cite{Klein:1999gv,Xing:2020hwh}. Secondly, one can also take an alternative view on this as an example of Entanglement Enabled Intensity Interferometry (E$^2$I$^2$)~\cite{COTLER2021168346} between two non-identical particles ($\pi^+$ and $\pi^-$). In fact, the components in the mass range of the $\rho^0$ consist of the $\rho^0$ particle and direct $\pi^+\pi^-$~\cite{star_collaboration_coherent_2017,theh1collaborationDiffractiveElectroproductionRho2010,alicecollaborationCoherentPhotoproductionRho2020a} that are subject to the same interference effect. In the example of Ref.~\cite{COTLER2021168346}, the sources of $\bf 1$ and $\bf 2$ are the two gold nuclei each emitting a pair of $\pi^+\pi^-$ and detectors $\bf A$ and $\bf B$ measure either $\pi^+$ or $\pi^-$. Due to the entanglement of the $\pi^+\pi^-$ at the source, there is a non-trivial interference term as shown in Eq.~4 of Ref.\cite{COTLER2021168346}. Finally, there exists a third scenario in which the initial $\rho^0$ wave functions are locked in phase through phase entanglements of the initial photons and Pomerons. In this case, the interference would only appear in the coherent process and would not produce any interference from the incoherent process~\cite{Zha:2018jin,ZhaPhysRevD.103.033007}. In all three cases, the interference occurs over the characteristic distance of the average impact parameter of the collisions, about 20 ${\rm fm}$, while the lifetime of the $\rho^0$ is only about 1 ${\rm fm}$. The decay daughter pions are spin zero particles. Our measurements of non-zero spin alignment ($[29.2\pm0.4({\rm stat.})\pm0.4({\rm syst.})]\%$ in Au+Au and $[23.8\pm0.6({\rm stat.})\pm0.6({\rm syst.})]\%$ in U+U) show a definite interference effect due to the non-locality of the pion wave functions. Through this measurement, we can also set a limit on whether or not the wave functions experience decoherence due to the decay process or other activity in their vicinity. 
The prediction from Model I matches well with data while the prediction from Model II is about 20\% above the data as shown in Fig.~\ref{fig:fig_5}B. This implies the coherence is at least 80\%.  



\section*{Conclusion}


This measurement of photo-nuclear production in Au+Au and U+U collisions constitutes the first utilization of the interacting photon's transverse linear polarization recently demonstrated by STAR in measurements of the $\gamma\gamma\rightarrow e^+e^-$ process. We observe a significant $\cos2\phi$ modulation through the $\rho^0\rightarrow \pi^+\pi^-$ production channel. 
The observed amplitude and structure of the $\cos2\phi$ vs. $P_T$ distribution (in Au+Au and U+U collisions) is qualitatively consistent with theoretical calculations that include quantum interference effects due to the photon's transverse linear polarization. 
We demonstrate that the $\cos2\phi$ modulation is absent in p+Au collisions, which are expected to be free from interference effects since only one amplitude predominantly contributes in those interactions.
Independent of theoretical models, we quantify the interference patterns observed in Au+Au and U+U measurements by studying the polarization dependence of the $|t|$ distribution in two dimensions. 
We are able to remove the effects of photon transverse momentum and two-source interference to extract the nuclear radius of gold and uranium. 
The resulting radii are systematically larger than the measured nuclear charge radii at lower energies. 
Furthermore, demonstrating this spin-induced orbital angular momentum interferometry offers a new avenue for studying nuclear geometry and gluon distribution within large nuclei at a quantitative level.

\bibliography{scibib}

\bibliographystyle{Science}

\section*{Acknowledgments}
We thank the RHIC Operations Group and RCF at BNL, the NERSC Center at LBNL, and the Open Science Grid consortium for providing resources and support.  This work was supported in part by the Office of Nuclear Physics within the U.S. DOE Office of Science, the U.S. National Science Foundation, the Ministry of Education and Science of the Russian Federation, National Natural Science Foundation of China, Chinese Academy of Science, the Ministry of Science and Technology of China and the Chinese Ministry of Education, the National Research Foundation of Korea, Czech Science Foundation and Ministry of Education, Youth and Sports of the Czech Republic, Hungarian National Research, Development and Innovation Office, New National Excellency Programme of the Hungarian Ministry of Human Capacities, Department of Atomic Energy and Department of Science and Technology of the Government of India, the National Science Centre of Poland, the Ministry  of Science, Education and Sports of the Republic of Croatia, RosAtom of Russia and German Bundesministerium fur Bildung, Wissenschaft, Forschung and Technologie (BMBF) and the Helmholtz Association.

\section*{MATERIALS AND METHODS}

This measurement was conducted with the STAR experiment~\cite{STAR:2005gfr} at the Relativistic Heavy Ion Collider (RHIC) using data collected in 2010, 2011, 2012, and 2015. Data from Au+Au collisions at a center-of-mass energy per nucleon-nucleon pair ($\sqrt{s_{_{NN}}}$) of 200 GeV, amounting to 1530 $\pm$ 150 $\mu$b$^{-1}$, were collected in 2010 and 2011. Data from U+U collisions at a $\sqrt{s_{_{NN}}} = 193$ GeV, amounting to 270 $\pm$ 30 $\mu$b$^{-1}$, was collected in 2012. 
Finally, data from p+Au collisions at $\sqrt{s_{_{NN}}} = 200$ GeV, amounting to 35 $\pm$ 3.5 nb$^{-1}$, was collected in 2015.

These data sets employed a triggering system based on signals from several STAR detectors to select ultra-peripheral collisions that may contain $\pi^+\pi^-$ pairs decayed from the photo-nuclear production of a $\rho^0$ vector meson or from direct photo-nuclear production of $\pi^+\pi^-$ pairs~\cite{judd_evolution_2018}. For the Au+Au and U+U datasets, the trigger accomplishes this by selecting events with neutron emission resulting from a mutual Coulomb dissociation process~\cite{bertulani_physics_2005} using two Zero Degree Calorimeters (ZDCs) located about 18 meters along the beam line on either side of the interaction point. The trigger requires a coincidence between signals in the two ZDCs~\cite{judd_evolution_2018} on either side of the interaction point, to reject beam-background interactions in which a single beam interacts with the beam pipe or other material. Requiring a coincidence in the ZDCs also helps select interactions occurring at the center of STAR where its detectors have roughly uniform acceptance. Since the interaction probabilities for mutual Coulomb excitation and for the diffractive photoproduction processes are approximately independent of one another for a given impact parameter, the selection of mutual Coulomb dissociation events provides an effective technique for selecting these exclusive processes. Mutual Coulomb dissociation~\cite{bertulani_physics_2005,baltz_two-photon_2009} at RHIC is well modeled~\cite{Baltz:1998ex} with an uncertainty of $\pm5\%$ and the various ratios of the cross sections from the model are in agreement with the experimental results within $\pm7\%$ uncertainty~\cite{ZDC130xs2002}. In addition to the neutron emission from mutual Coulomb dissociation, the trigger system also requires signals in the mid-rapidity STAR detectors (at large angles with respect to the beam) to be consistent with the presence of at least 2, but not more than 6 charged particles in order to effectively select exclusive events. Finally, in order to ensure that the selected events are diffractive in nature, the trigger system includes a veto on activity in the forward/backward regions between mid-rapidity and the ZDCs.  
Since a mutual Coulomb dissociation trigger cannot be used for p+Au collisions, a similar UPC trigger is implemented by employing Roman Pots to detect the minimally deflected proton~\cite{STARRP:2020dzd}. In this way candidate exclusive events, such as diffractive photo-nuclear interactions, can be efficiently triggered even in p+Au events that do not satisfy the mutual Coulomb dissociation trigger. 

\subsubsection*{Signal Selection}
In total $\sim$41$\times10^6$ Au+Au collisions, $\sim$23$\times10^6$ U+U collisions, and $\sim$1.8$\times10^8$ p+Au collisions were recorded using the trigger system. From the triggered events, candidate exclusive diffractive events were further required to: fall within $\pm$ 100 cm of the center of STAR, contain an interaction vertex formed from exactly 2 oppositely charged particles, and have no more than 2 additional background charged particles elsewhere in STAR. The reconstruction and identification of $\pi^+\pi^-$ pairs is accomplished using the Time Projection Chamber (TPC)~\cite{anderson_star_2003} and the Time of Flight (TOF) detectors~\cite{llope_tofppvpd_2004}. 
By measuring the track curvature, the TPC is capable of measuring the transverse momenta ($p_T$) of charged particles bent by a uniform 0.5 Tesla solenoidal magnetic field. The TPC has acceptance for $p_T > $100~MeV with a resolution $\sigma_{p_T}/p_{T}$ better than 1\% for tracks with momenta below 1~GeV.  The 3D tracks reconstructed by the TPC are required to have at least 15 hit points out of a possible maximum of 45 to ensure good quality.
In addition to tracking information, the TPC provides particle identification information via measurement of the mean specific ionization energy loss per unit length $\left(\langle dE/dx \rangle\right)$.
The measured $\langle dE/dx \rangle$ can be expressed in terms of the measurement resolution as $n\sigma$, where $n\sigma$ is the number of standard deviations from the expected $\langle dE/dx \rangle$ for a pion.  
The $n\sigma$ measurement of each track is used to construct the $\chi^2_{\pi\pi}$ for a $\pi^+\pi^-$ pair hypothesis, where $\chi^2_{\pi\pi} = n\sigma_1^2 + n\sigma_2^2$ and the subscripts 1,2 denote the two tracks in the pair.  

While the $\chi^2_{\pi\pi}$ variable alone can be used to select a relatively pure set of $\pi^+\pi^-$ pairs, those pairs may suffer from a small amount of contamination from $e^+e^-$, $p\bar{p}$, and $K^+K^-$ pairs when the daughter tracks have momenta above 200 MeV.
In order to further reject these possible types of contamination, the Time-of-Flight (TOF) detector, which provides precise timing measurement of charged particles with a time resolution of $\sim75$ ps can be used. 
The TOF identification technique is based on the measured time difference between the two tracks in a pair ($\Delta \mathrm{TOF} = t_2 - t_1$). Since the TPC provides measurement of the daughter particle momenta ($p_1, p_2$) and path lengths ($s_1, s_2$), the expected time difference for a $\pi^+\pi^-$ pair can be expressed as $\Delta \mathrm{TOF}^{E} = t_2^{E} - t_1^{E}$ where $t_{1,2}^{E} = \sqrt{ s_{1,2}^2 / c^{2} \times ( 1 + m_\pi^2/p_{1,2}^2 ) }$ (where $m_\pi$ is the mass of the charged pion $\approx139$ MeV/$c^2$). 
Finally, the double difference, $\Delta\Delta \mathrm{TOF} = \Delta \mathrm{TOF} - \Delta \mathrm{TOF}^{E}$, can be used to select $\pi^+\pi^-$ pairs while also rejecting contamination from $e^+e^-$, $p\bar{p}$, and $K^+K^-$ pairs.  Finally, for the Au+Au and U+U datasets, candidate $\pi^+\pi$ pairs are selected from those pairs with $\chi^2_{\pi\pi} < 8$ and $\Delta\Delta$TOF $<$ 750 ps. Since the TOF information was not available in the p+Au data set, only the $\chi^2_{\pi\pi} < 8$ condition is required. The effect of potential contamination from $e^+e^-$, $p\bar{p}$, and $K^+K^-$ pairs in the p+Au data set is found to be negligible and is included as a systematic uncertainty.


\subsection*{Detector Acceptance and Efficiency Corrections}
Instead of fitting to the $\phi$ distribution to extract the modulation strength, the $\cos2\phi$ distribution can be directly measured with respect to the various pair kinematics by taking the moments of the distribution. 
The strength of the modulation can then be determined directly as $2\times \langle \cos 2 \phi \rangle$, without the need for a fit, where the $\langle...\rangle$ indicate the average value defined as:
\begin{equation}
    2\langle \cos 2\phi \rangle_\mathrm{measured} = 2 \frac{\int_{-\pi}^{\pi} (1+\alpha(\phi) w(\phi) \cos2\phi)\cos 2\phi\ d\phi}{ \int_{-\pi}^{\pi} (1+\alpha(\phi) w(\phi) \cos2\phi) d\phi } 
    \label{eq:averagew}
\end{equation}
with $\alpha(\phi)$ being the true modulation strength and $w(\phi)$ the bias caused by imperfect detector efficiency and incomplete acceptance. Without loss of generality, both the signal and the detector effects can be expressed in terms of Fourier series:
\begin{equation}
    \begin{split}
    w(\phi) &= 1 + w_1 \cos\phi + w_2 \cos 2\phi + w_3 \cos 3\phi\ +\  ... \\
    \alpha(\phi) &= 1 + \alpha_1 \cos\phi + \alpha_2 \cos 2\phi + \alpha_3 \cos 3\phi\ +\  ...
    \label{eq:fourierdef}
    \end{split}
\end{equation}
In general the $\alpha$ and $w$ are also functions of the pair kinematics, e.g. $\alpha = \alpha(\phi, p_T, ..)$.
For an ideal detector there would be no bias (i.e. $w(\phi) = 1$) and the measured distribution would directly reflect the true distribution $\alpha(\phi)$.
Even though the STAR detector has full $2\pi$ coverage in azimuth, the limited acceptance in $p_T$ ($p_T > 200$ MeV) leads to a bias ($w(\phi) \neq 1$) and therefore, the measured $\phi$ distribution must be corrected to remove this detector effect. 
This acceptance bias is estimated directly from the data by constructing pairs with $\pi^+$ and $\pi^-$ tracks from different events (so-called mixed-event method). 
In this way the detailed acceptance and efficiency of the STAR detector can be reproduced to determine the acceptance effects with high precision.
Using this method, the bias on the $\cos2\phi$ modulation ($w_2$) is estimated and found to be small ($\ll 1\%$) for small pair $P_T$ but grows larger with increasing $P_T$ to a value of $\approx-10\%$ at a $P_T$ of 200 MeV. 
In principle, the detector bias can also lead to mixing between the various Fourier components. 
On the other hand, in practice this is not an issue, since all other components ($w_1, w_3, w_4,$ ...) are found to be consistent with zero in the $P_T$ range of interest. 
In that case, Eq.~\ref{eq:averagew} can be evaluated by substituting in the definitions of $\alpha$ and $w$ from Eq.~\ref{eq:fourierdef} and setting all $w_n$ to zero except for $n=2$. Evaluating and solving for the true signal yields:

\begin{equation}
    \alpha_2 = \frac{-2(\gamma_2 - w_2)}{-2 + \gamma_2 w_2}
\end{equation}

where $\gamma_2$ is the measured distribution, i.e $\gamma_2 = 2\langle \cos 2\phi \rangle_\mathrm{measured}$. 

\subsubsection*{Systematic Uncertainties}
In addition to the statistical uncertainties reported in Fig.~\ref{fig:fig_2} and Fig.~\ref{fig:fig_3}, we also evaluate possible sources of systematic bias for the $\phi$ measurement. The dominant sources of systematic uncertainty are: track selection and quality, pion identification, pion to muon decay, detector efficiency effects, and kinematic acceptance corrections. At low $P_T$ all of these sources are found to be smaller than the statistical uncertainty leading to point-to-point variations at the level of $3\%$ or less. 
At higher $P_T$ the pion to muon decay process is problematic because it potentially produces a self-correlation between $(\vec{p_1} + \vec{p_2})$ and $(\vec{p_1} - \vec{p_2})$. 
Since the muon tracks have approximately the correct azimuthal angle, but are reconstructed with the wrong mass (charged pion mass instead of muon mass), the pairs with one or more muons result in a $\phi$ distribution roughly peaked at $0$ and $\pm\pi$. 
This peak structure appears as anomalous components in a Fourier expansion of the distribution. 
The effect of muon contamination is estimated using a GEANT3-based simulation of $\pi^+\pi^-$ pairs through the STAR detector. 
In the measured mass range, the maximum effect is $\approx2\%$ for $P_T > 60$ MeV and is subtracted from the measured signal with the possible variation in the muon decay contribution taken as a systematic uncertainty. The other sources of uncertainty all contribute approximately equally for $P_T > 60$ MeV leading to a combined one sigma uncertainty on the amplitude of the $\cos2\phi$ modulation of $\sim3-6\%$. 

To study the systematical uncertainty of the choice of a form factor to extract the nuclear radius, we perform another analysis by dividing the $|t|$ distribution into different $\phi$ bins. In each $\phi$ bin, the $|t|$ distribution is fit with  
\begin{equation}
    f(t)=A_{c}\exp{(-b_Tt)}+\frac{A_{i}/Q_0^2}{(1+t/Q_0^2)^2},
    \label{eq:fit_expo}
\end{equation}
where $A_{c}$ is the amount of coherent production with an exponential shape and $A_{i}$ is the amount of incoherent production characterized by a dipole form factor. The value of $Q_0^2$ is fixed to $Q_0^2=0.099$ ${\rm GeV}^{-2}$ from previous measurement of the incoherent contribution~\cite{star_collaboration_coherent_2017}. The resulting $b_T$ as a function of $\phi$ is fitted with an empirical function to extract the nuclear radius at $\phi=\pi/2$. Using an empirical second order fit function of: 
\begin{equation}
    4/b_T=R_0^2+{\frac{1}{\epsilon_p}}\sigma_b^2(1+\epsilon_p\cos{2\phi}),
\end{equation}
we obtained $R_0=6.72\pm0.02$ fm for Au+Au and $R_0=7.37\pm0.03$ fm for U+U, and corresponding 
$\sigma_b=2.90\pm 0.05$ fm and $2.9\pm0.1$ fm.
The fit to the Woods-Saxon form factor is superior because 1) a much larger fit range is possible (which includes the second diffractive peak), and 2) Eq.~\ref{eq:fit_ws} describes the non-Gaussian shape better than Eq.~\ref{eq:fit_expo}. 
For these reasons, the exponential shape is not used to characterize the coherent contribution but as an estimate of the systematical uncertainty ($\pm0.05$ fm) for the form factor choice.

\subsection*{Tabulated Results}
We compile all the results from the measurements on radius in Table ~\ref{tab:STARRadius} and $\cos{2\phi}$ to compare with models and nuclear charge radius in Table ~\ref{tab:final}. 
\begin{table}
    \centering
    \begin{tabular}{|c|c|c|c|c|c|c|}
    \hline
          & $R_{inc}$ (fm) & $R(\phi=0^\circ)$ &  $R(\phi=90^\circ)$& fitted $R_0$ (Eq.~4) & Fitted $R_0$ (Eq.~9)& final \\ \hline
         Au &$7.47\pm0.02$&$7.86\pm0.03$&$7.15\pm0.03$&$6.62\pm0.03$&$6.72\pm0.02$&$6.53\pm0.03$\\
         U &$7.98\pm0.03$&$8.12\pm0.06$&$7.60\pm0.06$&$7.37\pm0.06$&$7.37\pm0.03$&$7.29\pm0.06$\\\hline
         \end{tabular}
    \caption{Results of the extracted radius from various methods and intermediate steps.}
    \label{tab:STARRadius}
\end{table}
\begin{table}
\centering
\begin{tabular}{|c|c|c|}
\hline
& $^{197}{\rm Au}$ &$^{238}{\rm U}$ \\ \hline
STAR $R$ (fm)&$6.53\pm0.03\pm0.05$&$7.29\pm0.06\pm0.05$\\
STAR $\langle\cos{2\phi}\rangle~(\%)$&$29.2\pm0.4({\rm stat.})\pm0.4({\rm syst.})$&$23.7\pm0.6({\rm stat.})\pm0.6({\rm syst.})$\\
$R_p$ (fm)& 6.38&6.87 \\
Model I (II) $R$ (fm)&6.38 (6.9)&\\\hline
\end{tabular}
\caption{Radius and $\cos{2\phi}$ from STAR data and those used or predicted in the models and nuclear charge radius ($R_p$). The reported $\langle\cos{2\phi}\rangle$ corresponds to $\pi^+\pi^-$ pairs with $0.65 < M_{\pi\pi} < 0.9$ GeV and $P_T < 0.06$ GeV.}
    \label{tab:final}
\end{table}

\subsubsection*{Woods-Saxon Form Factor Calculation}
As mentioned, in several cases the $dN/d|t|$ distributions are fit to Eq.~\ref{eq:fit_ws}, which characterizes the coherent contribution via the form factor of a Woods-Saxon distribution. In general, the form factor can be computed as the Fourier transform of the density distribution $(\rho_A)$:
\begin{equation}
    \label{eq:general_ff}
    F(k^2) = \frac{1}{Z e} \int d^3 r e^{i\vec{k} \cdot \vec{r} } \rho_A( \vec{r} ),
\end{equation}
where $Z$ is the charge of the nucleus, $e$ is the fundamental unit of charge, $\vec{k}$ is the momentum transfer and $\vec{r}$ is the position vector. For a spherically symmetric Woods-Saxon distribution, $\rho_A$ is a function of the radial distance only, and is defined as:
\begin{equation}
    \label{eq:ws}
    \rho_A(R) = \frac{ \rho_0 }{ 1 + \exp[ (R - R_A)/a ] },
\end{equation}
where $R$ is the radial distance from the center of the distribution, $R_A$ is the nuclear radius and $a$ is the surface thickness of the distribution. The factor $\rho_0$ is a normalization factor defined as $\rho_0 = 3A/(4\pi R_A^3)$, with $A$ being the atomic mass number for the given nucleus. In the case of a spherically symmetric distribution, Eq.~\ref{eq:general_ff} can be further simplified to:
\begin{equation}
    F(k^2) = \frac{4\pi\hbar}{Zek}\int R \rho_A(R) \sin{\left( kR/\hbar\right)} dR.
\end{equation}
Since Eq.~\ref{eq:ws} does not have an analytic Fourier transform, some authors choose to use an approximately equivalent analytic formula (See for instance Eq.~10 of Ref.~\cite{klein_starlight_2017}). Alternatively, the form factor for the Woods-Saxon distribution can be computed numerically, as is done in this analysis. We use the QAG adaptive Monte-Carlo integration technique~\cite{galassiGNUScientificLibrary2009} for computing the form factor, ensuring that the numerical accuracy is better than $0.1\%$.

\author{
M.~S.~Abdallah$^{5}$,
B.~E.~Aboona$^{57}$,
J.~Adam$^{7}$,
L.~Adamczyk$^{2}$,
J.~R.~Adams$^{41}$,
J.~K.~Adkins$^{32}$,
G.~Agakishiev$^{30}$,
I.~Aggarwal$^{43}$,
M.~M.~Aggarwal$^{43}$,
Z.~Ahammed$^{63}$
% ...
}

\normalsize{\rm{(STAR Collaboration)}}\\
\normalsize{$^{1}$Abilene Christian University, Abilene, Texas   79699}\\
\normalsize{$^{2}$AGH University of Science and Technology, FPACS, Cracow 30-059, Poland}\\
\normalsize{$^{3}$Alikhanov Institute for Theoretical and Experimental Physics NRC "Kurchatov Institute", Moscow 117218, Russia}\\
\normalsize{$^{4}$Argonne National Laboratory, Argonne, Illinois 60439}\\
\normalsize{$^{5}$American University of Cairo, New Cairo 11835, New Cairo, Egypt}\\
\normalsize{$^{6}$Ball State University, Muncie, Indiana, 47306, United States}\\
\normalsize{$^{7}$Brookhaven National Laboratory, Upton, New York 11973}\\
\normalsize{$^{8}$University of Calabria \& INFN-Cosenza, Italy}\\
\normalsize{$^{9}$University of California, Berkeley, California 94720}\\
\normalsize{$^{10}$University of California, Davis, California 95616}\\
\normalsize{$^{11}$University of California, Los Angeles, California 90095}\\
\normalsize{$^{12}$University of California, Riverside, California 92521}\\
\normalsize{$^{13}$Central China Normal University, Wuhan, Hubei 430079 }\\
\normalsize{$^{14}$University of Illinois at Chicago, Chicago, Illinois 60607}\\
\normalsize{$^{15}$Creighton University, Omaha, Nebraska 68178}\\
\normalsize{$^{16}$Czech Technical University in Prague, FNSPE, Prague 115 19, Czech Republic}\\
\normalsize{$^{17}$Technische Universit\"at Darmstadt, Darmstadt 64289, Germany}\\
\normalsize{$^{18}$ELTE E\"otv\"os Lor\'and University, Budapest, Hungary H-1117}\\
\normalsize{$^{19}$Frankfurt Institute for Advanced Studies FIAS, Frankfurt 60438, Germany}\\
\normalsize{$^{20}$Fudan University, Shanghai, 200433 }\\
\normalsize{$^{21}$University of Heidelberg, Heidelberg 69120, Germany }\\
\normalsize{$^{22}$University of Houston, Houston, Texas 77204}\\
\normalsize{$^{23}$Huzhou University, Huzhou, Zhejiang  313000}\\
\normalsize{$^{24}$Indian Institute of Science Education and Research (IISER), Berhampur 760010 , India}\\
\normalsize{$^{25}$Indian Institute of Science Education and Research (IISER) Tirupati, Tirupati 517507, India}\\
\normalsize{$^{26}$Indian Institute Technology, Patna, Bihar 801106, India}\\
\normalsize{$^{27}$Indiana University, Bloomington, Indiana 47408}\\
\normalsize{$^{28}$Institute of Modern Physics, Chinese Academy of Sciences, Lanzhou, Gansu 730000 }\\
\normalsize{$^{29}$University of Jammu, Jammu 180001, India}\\
\normalsize{$^{30}$Joint Institute for Nuclear Research, Dubna 141 980, Russia}\\
\normalsize{$^{31}$Kent State University, Kent, Ohio 44242}\\
\normalsize{$^{32}$University of Kentucky, Lexington, Kentucky 40506-0055}\\
\normalsize{$^{33}$Lawrence Berkeley National Laboratory, Berkeley, California 94720}\\
\normalsize{$^{34}$Lehigh University, Bethlehem, Pennsylvania 18015}\\
\normalsize{$^{35}$Max-Planck-Institut f\"ur Physik, Munich 80805, Germany}\\
\normalsize{$^{36}$Michigan State University, East Lansing, Michigan 48824}\\
\normalsize{$^{37}$National Research Nuclear University MEPhI, Moscow 115409, Russia}\\
\normalsize{$^{38}$National Institute of Science Education and Research, HBNI, Jatni 752050, India}\\
\normalsize{$^{39}$National Cheng Kung University, Tainan 70101 }\\
\normalsize{$^{40}$Nuclear Physics Institute of the CAS, Rez 250 68, Czech Republic}\\
\normalsize{$^{41}$Ohio State University, Columbus, Ohio 43210}\\
\normalsize{$^{42}$Institute of Nuclear Physics PAN, Cracow 31-342, Poland}\\
\normalsize{$^{43}$Panjab University, Chandigarh 160014, India}\\
\normalsize{$^{44}$NRC "Kurchatov Institute", Institute of High Energy Physics, Protvino 142281, Russia}\\
\normalsize{$^{45}$Purdue University, West Lafayette, Indiana 47907}\\
\normalsize{$^{46}$Rice University, Houston, Texas 77251}\\
\normalsize{$^{47}$Rutgers University, Piscataway, New Jersey 08854}\\
\normalsize{$^{48}$Universidade de S\~ao Paulo, S\~ao Paulo, Brazil 05314-970}\\
\normalsize{$^{49}$University of Science and Technology of China, Hefei, Anhui 230026}\\
\normalsize{$^{50}$South China Normal University, Guangzhou, Guangdong 510631}\\
\normalsize{$^{51}$Shandong University, Qingdao, Shandong 266237}\\
\normalsize{$^{52}$Shanghai Institute of Applied Physics, Chinese Academy of Sciences, Shanghai 201800}\\
\normalsize{$^{53}$Southern Connecticut State University, New Haven, Connecticut 06515}\\
\normalsize{$^{54}$State University of New York, Stony Brook, New York 11794}\\
\normalsize{$^{55}$Instituto de Alta Investigaci\'on, Universidad de Tarapac\'a, Arica 1000000, Chile}\\
\normalsize{$^{56}$Temple University, Philadelphia, Pennsylvania 19122}\\
\normalsize{$^{57}$Texas A\&M University, College Station, Texas 77843}\\
\normalsize{$^{58}$University of Texas, Austin, Texas 78712}\\
\normalsize{$^{59}$Tsinghua University, Beijing 100084}\\
\normalsize{$^{60}$University of Tsukuba, Tsukuba, Ibaraki 305-8571, Japan}\\
\normalsize{$^{61}$United States Naval Academy, Annapolis, Maryland 21402}\\
\normalsize{$^{62}$Valparaiso University, Valparaiso, Indiana 46383}\\
\normalsize{$^{63}$Variable Energy Cyclotron Centre, Kolkata 700064, India}\\
\normalsize{$^{64}$Warsaw University of Technology, Warsaw 00-661, Poland}\\
\normalsize{$^{65}$Wayne State University, Detroit, Michigan 48201}\\
\normalsize{$^{66}$Yale University, New Haven, Connecticut 06520}\\
\normalsize{{$^{*}${\rm Deceased}}} 
\end{document}